\documentclass[reprint,amsmath,amssymb,aps,pra]{revtex4-2}
\usepackage{graphicx,color}
\usepackage{bm}
\usepackage{hyperref}
\def\vec#1{\bm{#1}}
\def\abs#1{\left\lvert#1\right\rvert}
\def\vev#1{\langle#1\rangle}

\let\erf\relax\DeclareMathOperator{\erf}{erf}

\let\sgn\relax\DeclareMathOperator{\sgn}{sgn}

\def\scalar#1#2{\langle#1\vert#2\rangle}

\begin{document}

\title{Complex scaling flows in the quench dynamics of interacting particles}
\author{Tilman Enss}
\author{Noel Cuadra Braatz}
\author{Giacomo Gori}
\affiliation{Institut f\"ur Theoretische Physik,
  Universit\"at Heidelberg, 69120 Heidelberg, Germany}
\date{\today}

\begin{abstract}
  Many-body systems driven out of equilibrium can exhibit scaling
  flows of the quantum state.  For a sudden quench to resonant
  interactions between particles we construct a new class of
  analytical scaling solutions for the time evolved wave function with
  a complex scale parameter.  These solutions determine the exact
  dynamical scaling of observables such as the pair correlation
  function, the contact and the fidelity.  We give explicit examples
  of the nonequilibrium dynamics for two trapped fermions or bosons
  quenched to unitarity, for ideal Bose polarons, and for resonantly
  interacting, Borromean three-body systems.  These solutions reveal
  universal scaling properties of interacting many-body systems that
  arise from the buildup of correlations at short times after the
  quench.
\end{abstract}
\maketitle


\section{Introduction}
\label{sec:intro}

The quantum dynamics of strongly correlated many-body systems can
often be described as fluid flow \cite{schaefer2009}.  Near
equilibrium, the slow relaxation of conserved charges and currents is
governed by hydrodynamics \cite{landauVI, smith1989}.  Remarkably,
even some situations far from equilibrium are well described by the
equations of fluid dynamics, for instance the fast hydrodynamization
observed in relativistic nuclear collisions \cite{romatschke2019}.
Advances in ultracold quantum gas experiments now provide a new
platform to explore far-from-equilibrium quantum dynamics in a
controlled setting in the lab.  In particular, recent experimental and
theoretical studies have focused on the quench dynamics when strong or
resonant interactions are suddenly switched on in bulk fermion
\cite{bardon2014, luciuk2017}, Fermi polaron \cite{knap2012,
  cetina2016}, bulk boson \cite{sykes2014, fletcher2017, eigen2018,
  sun2020high} and Bose polaron \cite{drescher2020, drescher2021,
  skou2021} systems.  Understanding the validity of fluid dynamics in
these strongly correlated systems far from equilibrium has a wider
impact for finding simpler effective descriptions of complex quantum
dynamics.

A quench to strong interaction in a many-body system is generally a
hard problem.  However, at short times the dynamics is dominated by
few-body correlations between nearby quantum particles
\cite{sykes2014, qi2021maximum}, and similarly for an impurity
quenched to strong interaction with a surrounding medium
\cite{knap2012, drescher2021}.  This universal short-time quantum
dynamics applies equally to larger systems before the many-body time
scale is reached.  For longer times, instead, collective many-body
excitations dominate and conformal symmetry can determine the
long-time asymptotics \cite{maki2022dynamics}.

In this work, we focus on the short-time dynamics in an extreme
out-of-equilibrium setting and study few particles quenched from a
noninteracting state to resonant contact interactions in a harmonic
trapping potential.  After the quench, an initially stationary quantum
state becomes a highly excited state of the new Hamiltonian and can be
represented as a large superposition of new eigenstates with a
complicated time evolution.  For two interacting particles, however,
these eigenstates are known and we find the analytical form of the
time evolved wave function.  From this solution we obtain the
dynamical scaling of observables, in particular the full pair
correlation function $g^{(2)}(r,t)$.  For contact interactions in
three dimensions it diverges as $g^{(2)}(r,t) = C(t)/(4\pi r)^2$ for
short distances $r$ between the particles \cite{tan2008energetics,
  tan2008large}.  Starting from an initially noninteracting state,
strong contact correlations build up linearly in time and the contact
scales as $C(t)\propto\abs{\sin \omega_0t}$ with trap frequency
$\omega_0$ \cite{sykes2014}.  We find that due to this short-distance
singularity, the fidelity has an anomalous time dependence
$1-\gamma \abs t^{3/2}$ for short times, as discussed in
Sec.~\ref{sec:quench}.

The main goal of this work is to construct a new class of analytical
quench solutions in order to reveal universal scaling dynamics,
generalizing the two-particle example above.  We explain in
Sec.~\ref{sec:scaling} that this new class of solutions for the global
wave function has an analytical scaling form reminiscent of fluid
flow.  As a simple example, a quantum harmonic oscillator with time
dependent trapping potential can be transformed into a new
time-varying coordinate system where the Hamiltonian is stationary
\cite{pitaevskii1997, werner2006unitary, gritsev2010}.  The well-known
solutions of the stationary harmonic oscillator can then be
transformed back to the original coordinates where the dynamical wave
function assumes a scaling form with a global scale parameter
$\lambda(t)>0$.  In this work we show that also an interaction quench,
which suddenly changes the Bethe-Peierls boundary condition of the
wave function at short distance, can be brought into such a scaling
form.  However, we find that the quenched wave function is stationary
in \emph{complex} space and time coordinates, and the nonequilibrium
quench dynamics in the original space-time coordinates is described by
a scaling flow with a complex scale parameter
$\lambda(t)\in\mathbb C$.  Intriguingly, the complex time coordinate
runs backward in real time, such that the quench evolution of an
initial positive-energy state is given by the complex scaling flow of
a \emph{negative} energy stationary state.  Expressing the quench
dynamics as a scale transformation of a \emph{single} stationary state
constitutes a dramatic simplification compared to the generic time
evolution of a highly excited state represented as a large
superposition of eigenstates.  This is reminiscent of the complex
scaling used to express a resonance not as an infinite superposition
but as a single state of complex energy \cite{balslev1971spectral,
  reed1978vol4, bach1998quantum}.  Our earlier explicit quench
solution is an example of such a complex scaling flow.

After this general construction we apply the new class of solutions in
Sec.~\ref{sec:appl} to quench dynamics in few- and many-body systems,
specifically to universal short-time scaling of observables, to
quenched impurities in a Bose-Einstein condensate and to resonant,
Borromean three-body systems.  We conclude in Sec.~\ref{sec:disc} and
discuss how strong few-body correlations constrain an effective fluid
description of the strongly correlated quantum gas.


\section{Interaction quench dynamics}
\label{sec:quench}

To set the stage we begin by deriving a new analytical solution of
quench dynamics in the traditional way, as a superposition of
eigenstates of the Hamiltonian after the quench.  In the next section
this solution will be re-derived as an instance of complex scaling
flows.

Consider two distinguishable particles in a three-dimensional (3D)
harmonic trapping potential $V(r) = (m/2)\omega_0^2r^2$ with trap
frequency $\omega_0$.  The particles of mass $m$ shall interact via an
attractive contact interaction and are described by the Hamiltonian
\begin{align}
  \label{eq:Ham}
  H = \frac{p_1^2}{2m} + \frac{p_2^2}{2m} + \frac m2\omega_0^2(r_1^2 +
  r_2^2) + g\delta_\text{reg}^{(3)}(\vec r_1-\vec r_2).
\end{align}
We recapitulate the spectrum and eigenstates found in
\cite{busch1998}; in the following we compute the nonequilibrium
dynamics after a change in interaction \cite{kerin2020two}, which has
similarities to the one-dimensional case \cite{kehrberger2018}.

The center-of-mass motion in \eqref{eq:Ham} decouples from the
relative motion, and the wave functions factorize as
$\Psi(\vec C,\vec r) = \psi^\text{cm}(\vec C) \psi^\text{rel}(\vec r)$
with center-of-mass coordinate $\vec C=(\vec r_1+\vec r_2)/2$ and
relative coordinate $\vec r=\vec r_1-\vec r_2$.  In three dimensions
the contact interaction needs to be regularized, and we choose the
Fermi pseudopotential
$\delta_\text{reg}^{(3)}(\vec r) = \delta^{(3)}(\vec r)
\partial_rr\dotsm$ of strength $g=4\pi\hbar^2a/m$, which is fully
characterized by the $s$-wave scattering length $a$.  The interaction
affects only the relative motion, and only the $l=0$ partial wave
component for a zero-range interaction.  The contact pseudopotential
then leads to the Bethe-Peierls boundary condition for the relative
radial $l=0$ wavefunction as $r\to0$,
\begin{align}
  \label{eq:bethe}
  \psi^\text{rel}(r) = A\Bigl(\frac1r-\frac1a\Bigr) + \mathcal O(r).
\end{align}
The eigenfunctions for generic $a$ are Whittaker functions
$W_{a,b}(x)$ which decay sufficiently for $r\to\infty$,
\begin{align}
  \label{eq:psirel}
  \psi_\nu^\text{rel}(r) = r^{-3/2} W_{E^\text{rel}/2,1/4}(r^2/\ell^2)
\end{align}
up to normalization.  We express lengths in units of the relative
oscillator length $\ell=\sqrt{\hbar/\mu\omega_0}$ for reduced mass
$\mu=m/2$ and energies in units of the oscillator energy
$\hbar\omega_0$.  The energy eigenvalues of relative motion are given
by $E_\nu^\text{rel} = 2\nu+3/2$, where $\nu$ denotes the non-integer
generalization of the principal quantum number of the harmonic
oscillator wave function.  The wave function \eqref{eq:psirel}
satisfies the boundary condition \eqref{eq:bethe} if $\nu$ is related
to the scattering length $a$ as \cite{busch1998}
\begin{align}
  \label{eq:nu}
  \frac{\Gamma(-\nu)}{\Gamma(-\nu-\frac12)}
  = \frac{\Gamma(-\frac{E^\text{rel}}2+\frac34)}
  {\Gamma(-\frac{E^\text{rel}}2+\frac14)}
  = \frac\ell{2a}.
\end{align}
In the weakly interacting limit $a\to0^-$ with integer
$\nu=n=0,1,2,\dotsc$ one recovers the spectrum $E_{n,a\to0^-}=2n+3/2$
of the breathing modes of the noninteracting harmonic oscillator.  For
larger values of $1/a$ the energy levels decrease monotonically.  A
particularly interesting case is resonant scattering at $1/a=0$, where
the scattering amplitude reaches the maximum value consistent with
unitarity and scale invariance is restored.  At resonance (unitarity)
$\nu=n-1/2$ takes half-integer values for $n=0,1,2,\dotsc$ and the
eigenfunctions simplify to
\begin{align}
  \label{eq:psireluni}
  \psi_{n,1/a=0}^\text{rel}(r) =
  \frac{e^{-r^2/2\ell^2}H_{2n}(r/\ell)}
  {\pi^{3/4}2^n\sqrt{2(2n)!\ell}\, r}
\end{align}
with Hermite polynomials $H_n(x)$.  The associated energy eigenvalues
at resonance,
\begin{align}
  \label{eq:Ereluni}
  E_{n,1/a=0}^\text{rel} = 2n+1/2,
\end{align}
are again \emph{equally spaced} as in the noninteracting case, but
shifted downward by one unit of $\hbar\omega_0$.  Due to scale
invariance at unitarity (the scattering length drops out as a length
scale), an SO(2,1) symmetry emerges that generates the spectrum
\eqref{eq:Ereluni} at equidistant spacing $2\hbar\omega_0$
\cite{pitaevskii1997, werner2006unitary, nishida2007nonrel,
  werner2012, moroz2012scale}.

The knowledge of eigenstates allows one to analytically compute the
time evolution of the quantum gas after a quench from an ideal gas to
unitarity.  For definiteness we prepare the system in the harmonic
oscillator ground state both for the center-of-mass and for the
relative coordinate at $a=0^-$,
\begin{align}
  \label{eq:psi0}
  \psi_{0,a=0^-}^\text{rel}(\vec r)
  = \frac{e^{-r^2/2\ell^2}}{\pi^{3/4}\ell^{3/2}}.
\end{align}
When the interaction is suddenly quenched to unitarity at time $t=0$,
the wave function is projected onto the new eigenbasis of relative
motion with coefficients
\begin{align}
  a_n = (\psi_{n,1/a=0}^\text{rel},\psi_{0,a=0^-}^\text{rel})
  = \frac{2^n}{\sqrt{2(2n)!}\, \Gamma(3/2-n)}.
\end{align}
Computing
$\psi^\text{rel}(r,t) = \sum_{n=0}^\infty a_n
e^{-itE_{n,1/a=0}^\text{rel}} \psi_{n,1/a=0}^\text{rel}(r)$ we find
the full relative wave function after the quench to resonance as
\begin{multline}
  \psi^\text{rel}(r,t)
  = \frac{e^{-3i\omega_0t/2}}{\pi^{5/4}\ell^{3/2}} \Bigl[
  \frac{\sqrt{e^{2i\omega_0t}-1}}{r/\ell}\; e^{i(r^2/2\ell^2)\cot
    \omega_0t} \\
  +\sqrt\pi\,e^{-r^2/2\ell^2}
  \erf\Bigl(\frac{r/\ell}{\sqrt{e^{2i\omega_0t}-1}}\Bigr)\Bigr].
  \label{eq:psiquenchrel}
\end{multline}
The quench leaves the initial energy
$E^\text{rel}=\sum_n (2n+1/2)\abs{a_n}^2 = 3/2$ unchanged.  This is
expected from the dynamic sweep theorem because the expectation value
of the contact operator vanishes in the Gaussian initial state, as we
shall find below \cite{tan2008energetics, tan2008large}.

\begin{figure}[t]
  \centering
  \includegraphics[width=\linewidth]{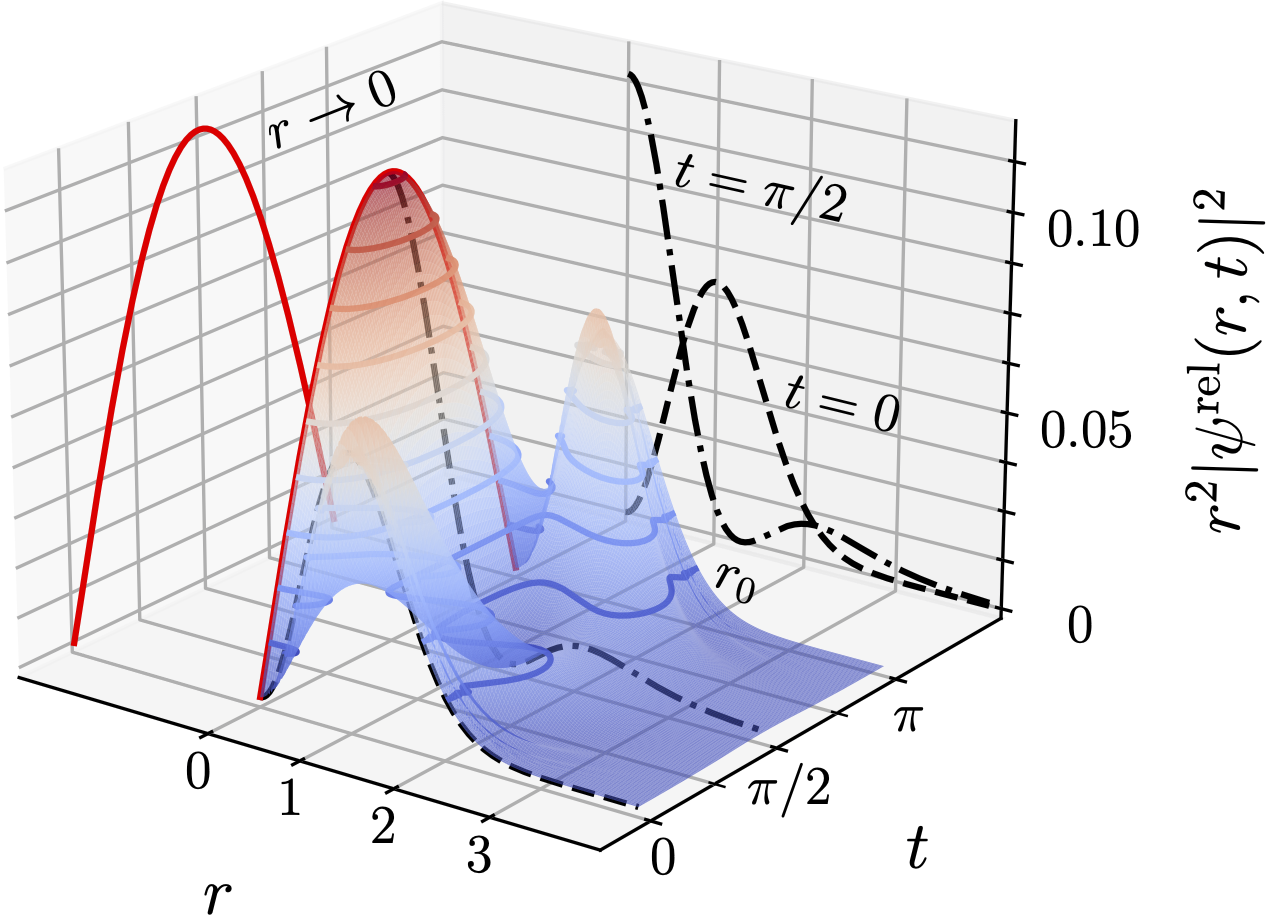}
  \caption{Pair correlation function
    $r^2g^{(2)}(r,t)=r^2\abs{\psi^\text{rel}(r,t)}^2$ vs distance $r$
    at time $t$ after a quench to unitarity ($1/a=0$), as given by
    Eq.~\eqref{eq:psiquenchrel}.  We give lengths in units of the trap
    length $\ell$ and times in units of $\omega_0^{-1}$.  The
    attractive interaction leads to a buildup of correlations at short
    distance $r\to0$.  At the same time, part of the correlation is
    pushed out so that an uncorrelated halo appears at $r=r_0$ and
    $\omega_0t=\pi/2$ (see text). At $r\to0$ one can read off the
    contact $C(t)$ (red solid line). Inside the trap the correlations
    are periodic in time with half the trap period. The black dashed
    and dash-dotted lines are the wavefunction at the beginning and
    halfway through the periodic motion, respectively.}
  \label{fig:paircorr}
\end{figure}

The pair correlation function of two particles at distance $r$ and
time $t$,
\begin{align}
  \label{eq:paircorr}
  g^{(2)}(r,t)=\abs{\psi^\text{rel}(r,t)}^2,
\end{align}
is shown in Fig.~\ref{fig:paircorr}.  The initial Gaussian profile
(weighted by $r^2$) is spread out over the trap length $r\simeq\ell$.
After the quench the attractive interaction pulls the particles
together at $r=0$ and correlations start to grow at short distance.
At short times, interference in the wave function
\eqref{eq:psiquenchrel} produces short-wavelength modulations that
scramble the correlation at all distances.  For longer times, however,
the correlation function becomes smooth again.  The harmonic
confinement brings the wave function back to its initial state at half
the trap period $\omega_0t=\pi$ and integer multiples; this is a
consequence of scale invariance and the SO(2,1) symmetry
\cite{werner2006unitary}.  Remarkably, at a quarter period
$\omega_0t=\pi/2$ the correlation develops a node and splits into two
disjoint regions at $r=r_0\approx1.306930\ell$, with the inner part
attracted toward $r=0$ by the contact interaction, while the outer
part is pushed farther out.  Note that during the whole time
evolution, the rms cloud size
$\vev{r^2}(t)=\int d^3r\, r^2\abs{\psi^\text{rel}(r)}^2 = (3/2)\ell^2$
remains constant even though the short-range correlations change
dramatically.

\begin{figure}[t]
  \centering
  \includegraphics[width=.85\linewidth]{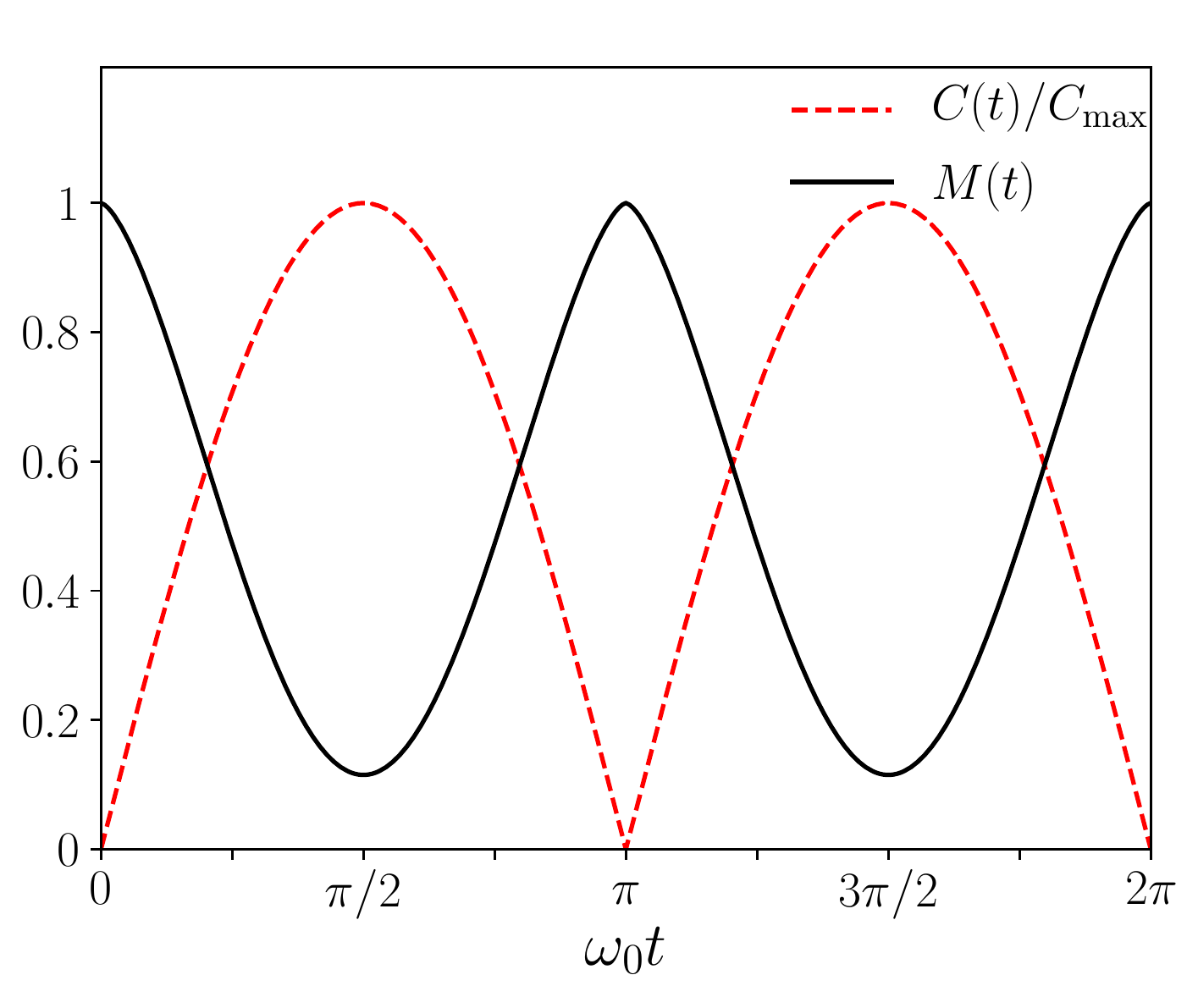}
  \caption{Contact $C(t)/C_\text{max}$ (dashed red line) at time $t$
    after a quench to unitarity ($1/a=0$), as given in
    Eq.~\eqref{eq:cquench}.  It rises linearly for short times and is
    periodic in time with half the trap period.  The fidelity or
    Loschmidt echo $M(t)$ (solid black line) in Eq.~\eqref{eq:fid} is
    nonanalytic as $1-\gamma \abs t^{3/2}$ for short times.}
  \label{fig:contact}
\end{figure}

In the short-distance limit the normalization \eqref{eq:bethe} of the
relative wave function is
\begin{align}
  \label{eq:Anorm}
  A(t) = \lim_{r\to0}r\psi^\text{rel}(r,t)
  = \frac{e^{-i\omega_0t}}{\pi^{5/4}\ell^{1/2}}\, \sqrt{2i\sin \omega_0t}.
\end{align}
This gives rise to the time evolution of the contact
\cite{tan2008energetics, tan2008large, sykes2014},
\begin{align}
  \label{eq:cquench}
  C(t)
  = \lim_{r\to0}(4\pi r)^2g^{(2)}(r,t)
  = \abs{4\pi A(t)}^2 
  = \frac{32}{\sqrt\pi\ell} \abs{\sin \omega_0t}.
\end{align}
As shown in Fig.~\ref{fig:contact}, immediately after the quench the
contact $C(t=0)=0$ vanishes as before the quench.  As the interaction
pulls together the particles, the contact grows linearly for short
times.  Eventually it reaches a maximum value of
$C_\text{max}=32/\sqrt\pi\ell$ at quarter period, before decreasing
again as a $\pi$-periodic function in time.

While the contact is sensitive to the wave function at short distance,
a measure of the evolution of the global quantum state is given by the
fidelity between initial and time evolved states, or Loschmidt echo
\cite{kehrberger2018, kerin2020two}
\begin{align}
  M(t)
  & = \abs{\scalar{\psi(0)}{\psi(t)}}^2 \notag \\
  & = \frac4{\pi^2}
    \abs{\sqrt{e^{2i\omega_0t}-1}+\arcsin(e^{-i\omega_0t})}^2.
  \label{eq:fid}
\end{align}
For short times the Loschmidt echo is anomalously suppressed as
$M(t) = 1-(8/3\pi)\abs{\omega_0t}^{3/2}+\mathcal O(t^{5/2})$ and
decreases faster than the usual $t^2$ behavior which follows at
intermediate times; this is due to the short-distance singularity of
the contact interaction.  The anomalously fast initial growth reflects
the fast scrambling of the wave function manifest in the ripples of
the pair correlation in Fig.~\ref{fig:paircorr} at short times.  The
same anomalous scaling is found for a quench in a bulk medium for
times shorter than the many-body times scale \cite{parish2016quantum}.
Eventually, the Loschmidt echo reaches a minimum of
$M(\omega_0t=\pi/2)=4[\sqrt2-\ln(1+\sqrt2)]^2/\pi^2\approx0.115$ at
quarter period and increases again to the next nonanalytic point at
$\omega_0t=\pi$, see Fig.~\ref{fig:contact}.


\section{Complex scaling flows}
\label{sec:scaling}

The exact quench dynamics in the harmonic trap \eqref{eq:psiquenchrel}
is a first example of a more general phenomenon and class of
analytical quench solutions.  In Eq.~\eqref{eq:psiquenchrel} the time
evolved state is written as a large superposition of eigenstates of
the quenched Hamiltonian which gives rise to a complicated transient
behavior.  We now show that the same wave function results from a
\emph{single stationary} state of the harmonic oscillator in a new
coordinate system that is related to the original one by a global
scale transformation.  This dramatically simplifies the description of
quench dynamics.


\subsection{Scaling dynamics}

Scaling flows are a powerful way to describe the nonequilibrium time
evolution of interacting quantum systems in the context of
hydrodynamics \cite{schaefer2009, schaefer2012scaling}, nonthermal
fixed points \cite{berges2008nonthermal} and trapped quantum gases
\cite{pitaevskii1997, werner2006unitary, murthy2019}.  As an example
consider a two-dimensional quantum harmonic oscillator with time
dependent trapping frequency $\omega(t)$:
\begin{align}
  \label{eq:harmosc}
  H(t) = \frac{\vec p^2}{2m} + \frac m2 \omega^2(t) \vec r^2.
\end{align}
Initially the harmonic oscillator shall have a constant trapping
frequency $\omega(t)=\omega_0$ and wave function
$\tilde\psi(\vec r,t)$.  If the trapping frequency $\omega(t)$ changes
in time for $t>0$, the wave function evolves as
\cite{werner2006unitary}
\begin{align}
  \label{eq:scalingform}
  \psi(\vec r,t) = \frac1{\lambda(t)}
  \exp\Bigl(\frac{im\vec r^2\dot\lambda(t)}{2\hbar\lambda(t)}\Bigr)
  \tilde\psi(\vec\rho,\tau).
\end{align}
The original wave function $\tilde\psi(\vec r,t)$ of the pre-quench
Hamiltonian is evaluated at new space and time coordinates
\begin{align}
  \label{eq:efftime}
  \vec r & \mapsto \vec\rho = \frac{\vec r}{\lambda(t)},
  & t & \mapsto \tau = \int_0^t \frac{dt'}{\lambda^2(t')}
\end{align}
in terms of a global, positive scale factor $\lambda(t) > 0$.  The
wave function $\psi(\vec r,t>0)$ after the quench
\eqref{eq:scalingform} satisfies the Schr\"odinger equation for given
time dependent trapping frequency $\omega(t>0)$ if the scale factor
evolves in time according to the Ermakov equation
\cite{pitaevskii1997, werner2006unitary, gritsev2010}
\begin{align}
  \label{eq:ermakov}
  \ddot\lambda+\omega^2(t)\lambda=\frac{\omega_0^2}{\lambda^3}
\end{align}
with initial conditions $\lambda(0)=1$, $\dot\lambda(0)=0$.  The
$1/\lambda$ term in \eqref{eq:scalingform} preserves the normalization
of the wave function in two-dimensional space, while the phase factor
that depends on space and time corresponds to a gauge transformation.
In case $\tilde\psi$ is a stationary state of the pre-quench
Hamiltonian at energy $E$, one can replace
$\tilde\psi(\vec\rho,\tau(t)) = e^{-iE\tau(t)/\hbar}
\tilde\psi(\vec\rho,0)$.

The time dependent coordinate transformation \eqref{eq:efftime} maps
the driven oscillator into a stationary one in new space $\vec\rho$
and time $\tau$ coordinates \cite{pitaevskii1997, werner2006unitary,
  gritsev2010}.  Solutions $\tilde\psi(\rho,\tau)$ of the
time-independent oscillator can then be transformed back to the
original coordinates $\vec r$, $t$ to yield the nonequilibrium scaling
solution \eqref{eq:scalingform}.  Remarkably, this scaling solution
for a single driven oscillator extends to interacting many-body
systems with scale invariant interactions that possess the SO(2,1)
symmetry in a harmonic trap, such as the unitary Fermi gas
\cite{werner2006unitary, nishida2007nonrel} or the 2D quantum gas
\cite{pitaevskii1997} up to the quantum scale anomaly
\cite{olshanii2010, hofmann2012, holten2018, peppler2018, murthy2019}.
It applies also to quantum statistical models where the quench
dynamics can be mapped to that of harmonic oscillators, such as the
spherical model \cite{syed2021dynamical}.

An interaction quench as discussed above in Sec.~\ref{sec:quench}
corresponds to a sudden change of the Bethe-Peierls boundary condition
\eqref{eq:bethe} for the relative $s$-wave function at $r=0$ from
noninteracting (scattering length $a=0^-$) to resonant interactions
($1/a=0$).  We demonstrate below that the ensuing quench dynamics is
again given by a scaling solution of the form \eqref{eq:scalingform}
but now with a \emph{complex} scale factor that solves the Ermakov
equation \eqref{eq:ermakov} with a different set of initial
conditions.  Remarkably, we find that the interaction quench dynamics
is obtained as the scaling in complex space and time of a
\emph{stationary} state $\tilde\psi$, albeit a different one from
before, which yields closed analytical expressions for the
nonequilibrium evolution.  We now derive this for the generalized case
of $N$ interacting particles in a harmonic trapping potential.
  

\subsection{Trapped N-particle systems}

Consider a three-dimensional $N$-particle system in a harmonic trap.
This is conveniently described in hyperspherical coordinates in terms
of center of mass $\vec C$, hyperradius $R$ and a collection of
hyperangles $\Omega$ \cite{blume2012few, werner2012}.  In the case of
scale invariant interaction the wave function factorizes as
\begin{align}
  \Psi_\text{trap}(X)
  = \psi^\text{cm}(\vec C) R^{-(3N-5)/2} F(R) \Phi(\Omega),
\end{align}
where $X=(\vec r_1, \dotsc, \vec r_N)$ is the vector of all particle
positions, $\vec C=(1/N)\sum_k \vec r_k$ denotes the center-of-mass
coordinate, $R=[\sum_k (\vec r_k-\vec C)^2]^{1/2}$ the hyperradius and
$\Omega$ the hyperangles.  The reason for choosing this coordinate
system is that an $N$-body interaction affects only the $R$ coordinate
and turns it into a one-dimensional problem which can be solved
analytically.  The hyperangular wave function satisfies the
Schr\"odinger equation
\begin{align}
  \left[ -\Delta_\Omega + \Bigl( \frac{3N-5}2 \Bigr)^2 \right]
  \Phi(\Omega) = s^2 \Phi(\Omega)
\end{align}
with Laplacian $\Delta_\Omega$ and energy eigenvalue
$s^2\in\mathbb R$.  Since the harmonic confinement affects only the
hyperradial and center-of-mass coordinates, the hyperangular solution
determines also the relative wave function in free space,
\begin{align}
  \Psi_\text{free}(X) = R^{s-(3N-5)/2} \Phi(\Omega).
\end{align}
For noninteracting particles in three dimensions, the ground state has
hyperangular eigenvalue $s=(3N-5)/2$ such that $s_{N=2} = 1/2$,
$s_{N=3} = 2$, etc.  For particles with resonant two-body interaction,
$s$ can take noninteger values, for instance $s=1.7727$ for
$N_\uparrow=2$, $N_\downarrow=1$ fermions \cite{blume2012few}.

Given the value $s$ of the hyperangular solution, the hyperradial wave
function is found by solving the 2D radial Schr\"odinger equation with
centrifugal barrier and oscillator confinement (from now on
$\hbar\equiv1$),
\begin{multline}
  \label{eq:Fschroed}
  -\frac{1}{2m} \left[ F''+\frac1R F' \right] + \left(
  \frac{s^2}{2mR^2} + \frac m2\omega_0^2R^2\right) F(R) \\
  = E_\text{rel} F(R),
\end{multline}
with the energy eigenvalue of relative motion $E_\text{rel}$ and
normalization $\int_0^\infty dR\, R\abs{F(R)}^2 = 1$.  For real $s$
there is a tower of universal states ($q\in\mathbb N_0$)
\begin{align}
  F_q(R)
  & = \sqrt{\frac{2(s+q)!}{q!s!^2}}\; \frac1R
    M_{E/2,s/2}(R^2) \\
  & = \sqrt{\frac{2q!}{(s+q)!}}\; R^s e^{-R^2/2} L_q^{(s)}(R^2)
\end{align}
with $R$ in units of the oscillator length
$L = \sqrt{\hbar/m\omega_0}$, while $L_q^{(s)}(x)$ denotes associated
Laguerre polynomials \cite{werner2012}.  The energy eigenvalues
$E_\text{rel} = (1+s+2q)\hbar\omega_0$ are equally spaced within each
tower of fixed $s$.  For positive $s>0$ and small hyperradius $R\to0$
the radial ground state wave function scales as
\begin{align}
  \label{eq:shortrange}
  F_0(R) \propto R^{s} (1+\mathcal O(R^2)).
\end{align}
Note that both the hyperangular and the hyperradial Schr\"odinger
equations depend only on $s^2$ and admit two solutions $s$, $-s$.
However, the sign change of $s$ selects a solution with a different
boundary condition for $R\to0$, namely $R^s$ vs.\ $R^{-s}$, and a
corresponding change of the ground-state energy from $E=1+s$ to
$E=1-s$.  This generalizes the Bethe-Peierls boundary condition
\eqref{eq:bethe} to $N$-body interaction with a condition on the
hyperradial wave function $F(R)$ for small $R\to0$ ($s>0$)
\cite{nishida2008universal, werner2012},
\begin{align}
  \label{eq:betheN}
  F(R)
  = A\Bigl( R^{-s} - \sgn(a)\frac{R^s}{\abs a^{2s}} \Bigr) + \dotsm.
\end{align}
The $R^s$ solution describes particles without $N$-body interaction
(scattering length $a\to0$) where $F(R)$ is bounded for $R\to0$.  The
$R^{-s}$ solution appears for $N$-body interaction of finite
scattering length $a\neq0$, and the $R^s$ part disappears completely
for resonant $N$-body interaction ($a\to\infty$).  For $-s\leq-1$, the
normalization of the radial function $F(R)$ can be formulated with a
short-distance cutoff that excludes the repulsive core as done for
$p$-wave and higher interaction \cite{pricoupenko2006modeling,
  pricoupenko2006pseudopotential}.

Consider now a quench from a noninteracting trapped $N$-particle Bose
gas to resonant $N$-body interactions.  This results in a Borromean
system with $N$-body but no $(N-1)$-body or smaller interaction
\cite{zhukov1993}.  Initially, the noninteracting gas has
$s=(3N-5)/2>0$, and the hyperradial ground state wave function
\begin{align}
  \label{eq:Finitial}
  F_0(R) = \sqrt{\frac2{s!}} \frac1R M_{(1+s)/2,s/2}(R^2)
  = \sqrt{\frac2{s!}} R^s e^{-R^2/2}
\end{align}
is normalizable and has energy $E_0=1+s=\frac32(N-1)$ for the relative
motion, which together with the center-of-mass energy
$E_\text{cm}=\frac32$ yields the total energy of
$E_\text{tot}=\frac32N$.  After the quench, the sign of $s$ is flipped
to $-s$ \cite{castin2012unitary} and the new resonant ground-state
energy becomes
\begin{align}
  E_0^\text{res} = 1-s.
\end{align}
The original wave function can be decomposed into a large
superposition of the new eigenstates $q\in\mathbb N_0$ with energies
$E_q^\text{res} = 1-s+2q$, and the interference between the tower
states results in the nonequilibrium quench dynamics as in
Eq.~\eqref{eq:psiquenchrel}, cf.~Fig.~\ref{fig:levelscheme}.


\subsection{Analytical quench solution}

\begin{figure}[!t]
  \centering
  \includegraphics[width=.9\linewidth]{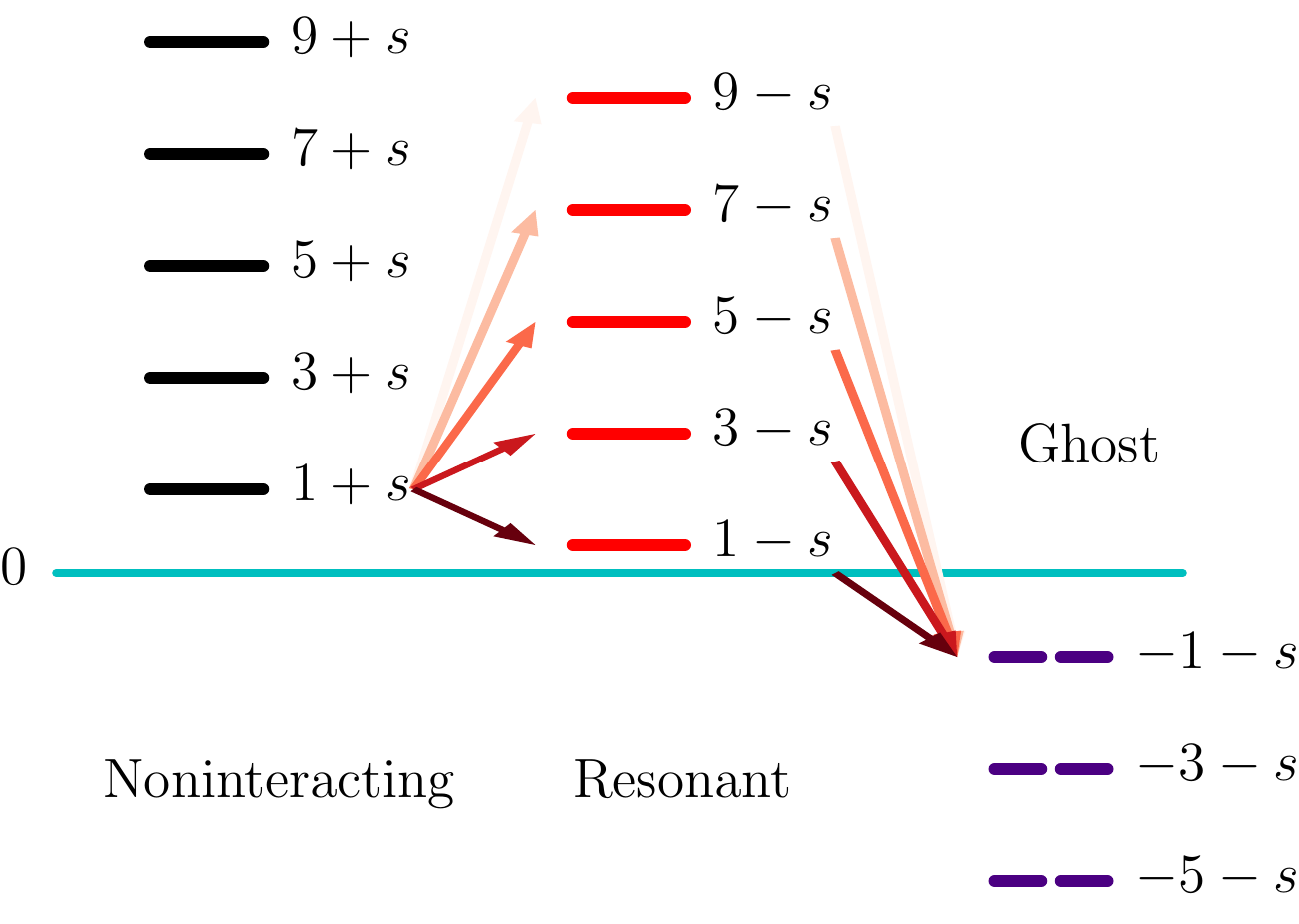}
  \caption{Complex scaling inverts the arrow of time: the
    noninteracting initial state of energy $1+s$ (in the
    ``Noninteracting'' tower) can be represented either as an infinite
    superposition of post-quench resonant states with energies
    $1-s+2q$, $q=0,1,2,\dotsc$ (the ``Resonant'' tower), or
    equivalently as the complex scaling of a \emph{single} stationary
    state of negative energy $-1-s$ (in the ``Ghost'' inverse tower)
    evolving backward in time.}
  \label{fig:levelscheme}
\end{figure}

We now construct the new analytical solution of $N$-body quench
dynamics.  We start with the general scaling form of the wave function
\eqref{eq:scalingform} with a \emph{complex} scale parameter
\begin{align}
  \label{eq:cscaling}
  \lambda(t)=\sqrt{e^{2i\omega_0t}-1}\in\mathbb C.
\end{align}
This solves the Ermakov equation for constant trapping frequency
$\omega_0$ but a vanishing scale parameter $\lambda(t\to0)=0$ at the
moment of the quench.  The scaling transformation \eqref{eq:efftime}
relates the original Hamiltonian to a stationary one in new complex
coordinates
\begin{align}
  \label{eq:tauscaling}
  \rho & = R\sqrt{-\frac{1+i\cot\omega_0t}2},
  & \tau
  & = \int_\varepsilon^t \frac{dt'}{\lambda^2(t')} 
    = -t-i\ln\frac{\lambda(t)}{\lambda_\varepsilon}
\end{align}
with
$\lambda_\varepsilon = \lambda(\varepsilon) \sim \varepsilon^{1/2}$
for short-time cutoff $\varepsilon\to0$.  We observe that the proper
time $\tau$ runs \emph{backwards} in real time $t$.  Hence, the
complex scaling inverts the energy of the stationary state
$\tilde\psi$.  Remarkably, there exists a \emph{negative-energy
  resonant state} with $q=-1$,
\begin{align}
  E_{-1}^\text{res} = 1-s-2 = -E_0,
\end{align}
whose energy is precisely the inverse of the pre-quench energy $E_0$.
Therefore, a \emph{single} stationary state is sufficient to describe
the full quench dynamics of the initial positive-energy state $E_0$
upon complex scaling.  Figure~\ref{fig:levelscheme} illustrates how
complex scaling maps the negative energy state to precisely match the
energy of the initial state.  The corresponding state $F_{-1}(\rho)$
results as the Whittaker function with negative second index $-s/2<0$,
\begin{align}
  \label{eq:Fminusone}
  F_{-1}^\text{res}(\rho)
  & = \mathcal N \frac{\lambda_\varepsilon^{1+s}}{\Gamma(1-s)}\,
    \frac1\rho M_{-(1+s)/2,-s/2}(\rho^2) \\
  & = \sqrt{\frac2{s!}}\lambda_\varepsilon^{1+s} \rho^s e^{\rho^2/2}
    [1-\Gamma(-s,\rho^2)/\Gamma(-s)]
\end{align}
with incomplete Gamma function $\Gamma(-s,x)$.  By construction,
$F_{-1}^\text{res}(R)$ is not normalizable for real $R\in\mathbb R$
because it is two levels below the oscillator ground state
$E_0^\text{res}$.  However, it becomes normalizable in complex space
coordinate $\rho = R/\lambda(t) \in \mathbb C$.  The complex scaling
solution \eqref{eq:scalingform} with the stationary state
$F_{-1}^\text{res}(\rho)$ then reads
\begin{align}
  \label{eq:Fscaling}
  F(R,t) = \frac{\exp(-iE_{-1}^\text{res}\tau)}{\lambda(t)}
  \exp\Bigl(\frac{iR^2\dot\lambda(t)}{2\lambda(t)}\Bigr)
  F_{-1}^\text{res}(\rho).
\end{align}
Since the proper time $\tau$ in \eqref{eq:tauscaling} runs backwards
in real time $t$ the phase factor from the time evolution of the
negative-energy stationary state $E_{-1}^\text{res}$ turns into that
of a positive-energy initial state $E_0$ (whose energy remains
unchanged by the quench) and an additional scale factor,
\begin{align}
  \label{eq:globalphase}
  e^{-iE_{-1}^\text{res}\tau}
  = e^{-iE_0t}
  \Bigl(\frac{\lambda(t)}{\lambda_\varepsilon}\Bigr)^{1+s} .
\end{align}
The $\lambda_\varepsilon$ term in the global phase
\eqref{eq:globalphase} compensates the corresponding term in the
normalization in \eqref{eq:Fminusone} to yield a finite result in the
$\varepsilon\to0$ limit.  Since $\lambda$ is complex, the gauge factor
changes not only the phase but also the amplitude, with complex
exponent
\begin{align}
  \frac{iR^2\dot\lambda}{2\lambda}
  = -\frac{R^2}4 (1-i\cot\omega_0t)
  = \frac12 \Bar\rho^2
\end{align}
in terms of the complex conjugate coordinate $\Bar\rho$.  Thus
$e^{(\Bar\rho^2+\rho^2)/2} = e^{-R^2/2}$ gives a pure amplitude factor
while $e^{(\Bar\rho^2-\rho^2)/2} = e^{iR^2\cot(\omega_0t)/2}$ is a
pure phase.  Collecting the terms in \eqref{eq:Fscaling}
we arrive at our main result, the analytical quench solution
\begin{align}
  \label{eq:Fscalingfinal}
  F(R,t)
  & = e^{-iE_0t} F_0(R)
    \Bigl[1-\frac{\Gamma\bigl(-s,\frac{R^2}{e^{2i\omega_0t}-1}\bigr)}
    {\Gamma(-s)} \Bigr].
\end{align}

It remains to be shown that the scaling solution \eqref{eq:Fscaling},
\eqref{eq:Fscalingfinal} satisfies (i) the Schr\"odinger equation
\eqref{eq:Fschroed} and (ii) the Bethe-Peierls boundary condition
$F(R,t)\sim R^{-s}$ for $t>0$, and (iii) is continuous with the
initial state \eqref{eq:Finitial} for $t\to0$.  In fact, (i) follows
because the complex scale factor \eqref{eq:cscaling} satisfies the
Ermakov equation \eqref{eq:ermakov} and $F_{-1}^\text{res}(R)$ is a
stationary solution of the Schr\"odinger equation, even though is has
negative energy and is not normalizable.  The boundary condition (ii)
for $t>0$ follows from the short-distance expansion of
\eqref{eq:Fscalingfinal} using
$\Gamma(-s,\rho^2\to0) = \rho^{-2s}/s+\dotsm$.  Continuity (iii)
requires $F(R,t\to0) = F_0(R)$: for short times
$\lambda^2\to2i\omega_0t$ and $\rho^2=R^2/\lambda^2 \to -i\infty$
becomes large. In the $t\to0$ limit the incomplete Gamma function
\begin{align}
  \label{eq:incomplete}
  \abs{\Gamma\Bigl(-s,\frac{R^2}{2it}\Bigr)}
  \sim \abs{\frac t{R^2}}^{1+s}
\end{align}
vanishes.  We thus obtain the continuity of the wave function for
short times,
\begin{align}
  \label{eq:continuity}
  F(R,t\to0)
  = \sqrt{\frac2{s!}} R^s\, e^{-R^2/2}
  = F_0(R).
\end{align}
Hence, the full initial wave function with noninteracting $R^s$
boundary condition is recovered for short times $t\to0$ or for
distances $R\gtrsim R_d(t)=\sqrt{2Dt}$ ($\abs\rho\gtrsim1$) larger
than the diffusion scale with quantum diffusivity $D\simeq\hbar/m$
\cite{enss2012spin}.  Complex scaling replaces this by the resonant
$R^{-s}$ boundary condition for longer times or shorter distances
$R\lesssim R_d(t)$.


\subsection{Example: two particles in a harmonic trap}
\label{sec:twobody}

The general analytic solution \eqref{eq:Fscalingfinal} recovers our
earlier result for $N=2$ particles derived in Sec.~\ref{sec:quench}.
Indeed, with $s=1/2$ one has $E_0=3/2$, $E_0^\text{res}=1/2$ and
$E_{-1}^\text{res} = -3/2 = -E_0$.  The post-quench stationary state
at negative energy is
\begin{align}
  \label{eq:Ftwobody}
  F_{-1}^\text{res}(\rho)
  & = \frac{2\lambda_\varepsilon^{3/2}}{\pi^{3/4}}\,
    \rho^{-1/2} e^{-\rho^2/2}[1+\sqrt\pi \rho\, e^{\rho^2} \erf(\rho)],
\end{align}
while the quench solution reads
\begin{align*}
  F(R,t)
  = e^{-iE_0t} \frac{2R^{1/2}}{\pi^{3/4}} \bigl[ \frac1\rho
  e^{i(R^2/2)\cot\omega_0t} + \sqrt\pi e^{-R^2/2} \erf\rho \bigr].
\end{align*}
We thus find the quenched 3D wave function
$\psi(r,t) = (4\pi r)^{-1/2} F(R/L=r/\ell,t)$ in agreement with
Eq.~\eqref{eq:psiquenchrel}.


\section{Applications}
\label{sec:appl}

From the analytical quench solution \eqref{eq:Fscalingfinal} it is now
straightforward to obtain the dynamical scaling of observables after
the quench.

\subsection{Dynamical scaling of observables}
\label{sec:obs}

While the energy of the initial state remains unchanged after the
quench, the density profile evolves in time.  In particular, the
$N$-body hyperradial correlation function ($s>0$)
\begin{align}
  g^{(N)}(R,t)
  & = \abs{F(R,t)}^2 
    = \frac2{s!}
    R^{2s}e^{-R^2}\abs{1-\frac{\Gamma(-s,\rho^2)}{\Gamma(-s)}}^2
    \notag \\
  & = C^{(N)}(t) R^{-2s} + O(R^{-2s+2})
  \label{eq:Ncorrelsing}
\end{align}
manifests how the overall scale of the gas responds to a change in the
$N$-body interaction.  This is the $N$-body generalization of the pair
correlation function \eqref{eq:paircorr} for $N=2$ and $s=1/2$.  
Figure~\ref{fig:paircorr} illustrates how after the quench the inner
part $R\lesssim L$ is pulled in, while the outer part $R\gtrsim L$ is
pushed further out.  Despite the internal motion, the average cloud
size (virial)
\begin{align}
  \vev{R^2}(t)
  & = \int_0^\infty dR\, R\, R^2\, g^{(N)}(R,t)
  = (1+s)L^2
\end{align}
remains constant after the quench for generic $s$, extending our
result for $s=1/2$ below Eq.~\eqref{eq:paircorr}.

In the short-distance limit the $N$-body hyperradial correlations
\eqref{eq:Ncorrelsing} are singular for resonant interaction as a
consequence of the Bethe-Peierls boundary condition \eqref{eq:betheN}.
The dynamical scaling of this singularity is given by the $N$-body
contact parameter
\begin{align}
  C^{(N)}(t)
  & = \lim_{R\to0}\abs{R^sF(R,t)}^2 
    = \frac2{s!s^2} \abs{\lambda(t)}^{4s} \notag \\
  & = \frac{2^{2s+1}}{s!s^2} \abs{\sin\omega_0t}^{2s}.
    \label{eq:contactN}
\end{align}
As discussed above, for $s\geq1$ a short-distance cutoff $R>R_c$ can
be used and the contact is read off just outside the cutoff radius.
The $N$-body contact is initially zero for an $N$-body noninteracting
state and rises as $\abs t^{2s}$ for short times to reach a maximum
value at quarter period $\omega_0t=\pi/2$.  This generalizes our
earlier result \eqref{eq:cquench} for the time dependent contact of
two particles with $s=1/2$.

Finally, the wave function overlap of the time evolved initial and
quenched states deviates
from unity as $t^{1+s}$ for short times,
\begin{align}
  \label{eq:fidN}
  \scalar{\psi_0(t)}{\psi(t)}
  & = e^{iE_0t} \int_0^\infty dR\, RF_0(R)F(R,t) \\
  & = 1-\frac{\lambda^{2(1+s)}{}_2F_1(1,1+s,2+s,-\lambda^2)}{(1+s)!(-1-s)!}
     \notag \\
  & = 1 - \frac{(2i\omega_0t)^{1+s}}{(1+s)!(-1-s)!} + \dotsm
\end{align}
with hypergeometric function ${}_2F_1(a,b,c,x)$.  For $s=1/2$ this
recovers the scaling of the Loschmidt echo for two particles
\eqref{eq:fid}.


\subsection{Many-body system in free space: \\
  Resonant impurity in an ideal BEC}

The quench dynamics of a harmonically trapped gas straightforwardly
includes the dynamics in free space by taking the limit of vanishing
trap frequency $\omega_0\to0$.  To see this, we include again units
and find
\begin{align}
  \rho^2 = \frac{R^2/(\hbar/m\omega_0)}{e^{2i\omega_0t}-1}
  \overset{\omega_0\to0}\longrightarrow
  \frac{mR^2}{2i\hbar t}
\end{align}
with $\lambda(t) \approx \sqrt{2i\omega_0t}$, while
$\exp(-R^2/2L^2)\to1$ in this limit.  The analytical quench dynamics
in free space is thus given by the wave function
\begin{align}
  \label{eq:Ffree}
  F_\text{free}(R,t) = R^s \Bigl(
  1-\frac{\Gamma(-s,mR^2/2i\hbar t)} {\Gamma(-s)} \Bigr).
\end{align}

The exact quench evolution \eqref{eq:Ffree} applies directly to an
ideal Bose-Einstein condensate (BEC) with a heavy impurity particle,
which is suddenly quenched to resonant interaction with the
surrounding condensate particles and thereby forms an ideal Bose
polaron \cite{drescher2021}.  The condensate wave function $\phi(r,t)$
at distance $r$ from the impurity agrees with the relative wave
function \eqref{eq:psiquenchrel}, \eqref{eq:Ffree} up to an overall
normalization factor for $N$ particles in the condensate, and we find
\begin{align}
  \label{eq:phitrap}
  \phi(r,t) = \sqrt N \psi^\text{rel}(r,t)
  \overset{\omega_0\to0}\longrightarrow
  \sqrt{\frac N{4\pi r}}\, F_\text{free}(r,t).
\end{align}
For a uniform BEC in the $\omega_0\to0$ limit we thus obtain the
quench solution
\begin{align}
  \label{eq:phiuniform}
  \phi(r,t)
  & = \lim_{\omega_0\to0} \sqrt{\pi^{3/2} n \ell^3}\,
    \psi^\text{rel}(r,t) \\
  & = \sqrt n\Bigl( \sqrt{\frac{2i\hbar t}{\pi mr^2}}\,
    e^{-m r^2/2i\hbar t}
    + \erf\sqrt{\frac{m r^2}{2i\hbar t}}\Bigr).
\end{align}
Here $m=m_\text{B}$ denotes the reduced mass between a boson of mass
$m_\text{B}$ and the infinitely heavy impurity.  This result
reproduces the exact quench solution for the ideal Bose polaron in a
uniform BEC derived recently in a continuum computation
\cite{drescher2021}.


\subsection{Borromean system with three-body interaction}
\label{sec:threebody}

To study quench dynamics beyond two particles we consider a bosonic
three-body system ($N=3$) which is initially noninteracting ($s=2$)
with relative ground-state energy $E_0=1+s=3$.  A quench of the
\emph{three}-body interaction imposes a sudden change of the $N$-body
Bethe-Peierls boundary condition \eqref{eq:betheN} on the hyperradial
wave function, while the two-body sector in the hyperangular part
$\Phi(\Omega)$ remains unaffected.  This creates a Borromean system
with three-body but no two-body interaction, which occurs both in
nuclei \cite{zhukov1993, hongo2022universal} and in ultracold gases
\cite{nishida2008universal, zwerger2019quantum}, for instance near a
three-body resonance \cite{fletcher2017}.

For the initially noninteracting gas with integer $s=2$ the Whittaker
$M$ function \eqref{eq:Fminusone} is undefined.  Instead, one can
write the quench solution as a linear combination of the two linearly
independent regular solutions $W_{(1+s)/2,s/2}(-\rho^2)$ and
$W_{-(1+s)/2,s/2}(\rho^2)$.  For noninteger $s$ the coefficients are
fixed by the initial condition $t\to0$ and the boundary condition
$R\to0$, and we recover \eqref{eq:Fscalingfinal}.  For integer $s=2$,
instead, we obtain the analytical quench solution
\begin{align}
  \label{eq:Fs2}
  F(R,t) = e^{-iE_0t} R^2 e^{-R^2/2} \Bigl[ 1
  + \frac{2i}\pi \Gamma(-2,\rho^2) \Bigr].
\end{align}
This wave function develops a node at intermediate distance at quarter
period $\omega_0t=\pi/2$, in analogy to the $s=1/2$ case above.
Following the discussion in Sec.~\ref{sec:obs}, we predict that the
three-body contact grows in time as $C^{(3)}(t) \sim t^4$.  This is
consistent with a recent experiment which found that three-body
correlations grow slower than two-body ones after an interaction
quench \cite{fletcher2017}.  If, instead, three bosons are already
resonantly interacting in the two-body sector with $s=4.465$
\cite{blume2012few} in the initial state before the quench, we expect
an anomalous growth law $C^{(3)}(t) \sim t^{8.93}$ reminiscent of
unparticle physics \cite{georgi2007, hammer2021unnuclear}.


\section{Discussion}
\label{sec:disc}

In conclusion, we have shown that $N$-particle systems quenched to
resonant $N$-body interaction exhibit scaling dynamics with a complex
scale factor, with explicit examples given for $N=2,3$.  This provides
a fully analytical form of the nonequilibrium dynamics as the complex
scaling of a single negative-energy stationary state.  The exact
few-body quench dynamics determines also the universal dynamics of a
many-body system at times $t\lesssim \hbar/E_F$ shorter than the
many-body time scale where medium effects become important
\cite{qi2021maximum}.

The complex scaling flow allows us to predict the dynamical scaling of
observables after the quench.  We find that the integrated two-body
contact \eqref{eq:cquench} grows linearly in time at short times after
a quench from an ideal to a unitary Fermi gas, with the growth rate
$C(t)\propto (\hbar n/m) t$ proportional to density \cite{sykes2014}.
This could be observed with state-of-the-art cold atom experiments
that measure the two- and three-body contact on very short time scales
\cite{bardon2014, luciuk2017, fletcher2017}.  In general, the $N$-body
contact scales universally as $C^{(N)}(t) \sim t^{2s}$ after the
quench, while the fidelity is anomalously suppressed as
$M(t) = 1-\gamma \abs t^{1+s}$.  For a three-body system where
resonant three-body interactions are switched on, this leads to a
characteristic scaling with $s=2$ (without two-body interaction) or
$s=4.465$ (resonant two-body interaction in $l=0$ state).  In our
discussion we assumed scale invariance and did not consider Efimov
three-body bound states with imaginary $s=1.00624i$ that break
continuous scale invariance and lead to modulations of the three-body
contact \cite{colussi2018dynamics, colussi2019bunching}.
Nevertheless, approaching the threshold for three-body bound states
provides a way to realize resonant three-body interactions in
experiment \cite{fletcher2017}.  Quenches into these states could be a
worthwhile topic for future study.

A different question is how an $N>2$ particle system evolves after a
quench in the two-body interacion.  In this case, the quench affects
also the hyper\emph{angular} part of the wave function, and the
nonequilibrium evolution might involve several towers of states with
the same total angular momentum but different values of $s$ for their
primary states \cite{blume2012few, bekassy2022}.

Such strong contact correlations have implications for the description
of fluid flow.  In general, transport can be described by the
Bogoliubov-Born-Green-Kirkwood-Yvon (BBGKY) hierarchy of particle
distribution functions where the evolution of the single-particle
distribution $f_1$ depends on the two-particle distribution $f_2$,
which in turn depends on higher distributions \cite{smith1989}.  In a
dilute gas, the property of molecular chaos means that particle
distributions are uncorrelated and one can set $f_2=f_1^2$: in this
way, the hierarchy of equations of motion closes and one can
explicitly compute the collision integral in the Boltzmann equation.
Our model system is very dilute with an interaction range
$\abs{r_e}\ll n^{-1/3}$ much shorter than the mean particle spacing;
at the same time, however, the strong contact correlations
$g^{(2)}(r,t)$ in Fig.~\ref{fig:paircorr} violate molecular chaos
$f_2\neq f_1^2$ and invalidate a Boltzmann approach formulated solely
in terms of the fermionic single-particle distribution but without
two-particle pair correlations.  Indeed, recent computations of the
bulk viscosity \cite{dusling2013, enss2019bulk, nishida2019,
  hofmann2020, fujii2020, maki2020role} and thermal conductivity
\cite{frank2020quantum} of strongly interacting Fermi gases reveal the
importance of contact correlations for transport in extension of the
fermionic Boltzmann formulation.  The initial buildup of few-body
correlations \cite{colussi2019bunching, musolino2022bose} should be
part of an effective fluid description of quench dynamics.


\begin{acknowledgments}
  We thank N.~Defenu, M.~Drescher, J.~Maki, J.~Thywissen, and
  W.~Zwerger for useful discussions.  This work is supported by the
  Deutsche Forschungsgemeinschaft (DFG, German Research Foundation),
  project-ID 273811115 (SFB1225 ISOQUANT) and under Germany’s
  Excellence Strategy EXC2181/1-390900948 (the Heidelberg STRUCTURES
  Excellence Cluster).
\end{acknowledgments}


\bibliography{all}

\begin{thebibliography}{62}%
\makeatletter
\providecommand \@ifxundefined [1]{%
 \@ifx{#1\undefined}
}%
\providecommand \@ifnum [1]{%
 \ifnum #1\expandafter \@firstoftwo
 \else \expandafter \@secondoftwo
 \fi
}%
\providecommand \@ifx [1]{%
 \ifx #1\expandafter \@firstoftwo
 \else \expandafter \@secondoftwo
 \fi
}%
\providecommand \natexlab [1]{#1}%
\providecommand \enquote  [1]{``#1''}%
\providecommand \bibnamefont  [1]{#1}%
\providecommand \bibfnamefont [1]{#1}%
\providecommand \citenamefont [1]{#1}%
\providecommand \href@noop [0]{\@secondoftwo}%
\providecommand \href [0]{\begingroup \@sanitize@url \@href}%
\providecommand \@href[1]{\@@startlink{#1}\@@href}%
\providecommand \@@href[1]{\endgroup#1\@@endlink}%
\providecommand \@sanitize@url [0]{\catcode `\\12\catcode `\$12\catcode
  `\&12\catcode `\#12\catcode `\^12\catcode `\_12\catcode `\%12\relax}%
\providecommand \@@startlink[1]{}%
\providecommand \@@endlink[0]{}%
\providecommand \url  [0]{\begingroup\@sanitize@url \@url }%
\providecommand \@url [1]{\endgroup\@href {#1}{\urlprefix }}%
\providecommand \urlprefix  [0]{URL }%
\providecommand \Eprint [0]{\href }%
\providecommand \doibase [0]{https://doi.org/}%
\providecommand \selectlanguage [0]{\@gobble}%
\providecommand \bibinfo  [0]{\@secondoftwo}%
\providecommand \bibfield  [0]{\@secondoftwo}%
\providecommand \translation [1]{[#1]}%
\providecommand \BibitemOpen [0]{}%
\providecommand \bibitemStop [0]{}%
\providecommand \bibitemNoStop [0]{.\EOS\space}%
\providecommand \EOS [0]{\spacefactor3000\relax}%
\providecommand \BibitemShut  [1]{\csname bibitem#1\endcsname}%
\let\auto@bib@innerbib\@empty
\bibitem [{\citenamefont {Sch{\"a}fer}\ and\ \citenamefont
  {Teaney}(2009)}]{schaefer2009}%
  \BibitemOpen
  \bibfield  {author} {\bibinfo {author} {\bibfnamefont {T.}~\bibnamefont
  {Sch{\"a}fer}}\ and\ \bibinfo {author} {\bibfnamefont {D.}~\bibnamefont
  {Teaney}},\ }\bibfield  {title} {\bibinfo {title} {{Nearly perfect fluidity:
  from cold atomic gases to hot quark gluon plasmas}},\ }\href
  {https://doi.org/10.1088/0034-4885/72/12/126001} {\bibfield  {journal}
  {\bibinfo  {journal} {Rep.\ Prog.\ Phys.}\ }\textbf {\bibinfo {volume}
  {72}},\ \bibinfo {pages} {126001} (\bibinfo {year} {2009})}\BibitemShut
  {NoStop}%
\bibitem [{\citenamefont {Landau}\ and\ \citenamefont
  {Lifshitz}(1987)}]{landauVI}%
  \BibitemOpen
  \bibfield  {author} {\bibinfo {author} {\bibfnamefont {L.~D.}\ \bibnamefont
  {Landau}}\ and\ \bibinfo {author} {\bibfnamefont {E.~M.}\ \bibnamefont
  {Lifshitz}},\ }\href@noop {} {\emph {\bibinfo {title} {{Fluid Mechanics}}}}\
  (\bibinfo  {publisher} {Butterworth-Heinemann},\ \bibinfo {address}
  {Oxford},\ \bibinfo {year} {1987})\BibitemShut {NoStop}%
\bibitem [{\citenamefont {Smith}\ and\ \citenamefont
  {Jensen}(1989)}]{smith1989}%
  \BibitemOpen
  \bibfield  {author} {\bibinfo {author} {\bibfnamefont {H.}~\bibnamefont
  {Smith}}\ and\ \bibinfo {author} {\bibfnamefont {H.~H.}\ \bibnamefont
  {Jensen}},\ }\href@noop {} {\emph {\bibinfo {title} {{Transport
  Phenomena}}}}\ (\bibinfo  {publisher} {Oxford University Press},\ \bibinfo
  {address} {Oxford, UK},\ \bibinfo {year} {1989})\BibitemShut {NoStop}%
\bibitem [{\citenamefont {Romatschke}\ and\ \citenamefont
  {Romatschke}(2019)}]{romatschke2019}%
  \BibitemOpen
  \bibfield  {author} {\bibinfo {author} {\bibfnamefont {P.}~\bibnamefont
  {Romatschke}}\ and\ \bibinfo {author} {\bibfnamefont {U.}~\bibnamefont
  {Romatschke}},\ }\href@noop {} {\emph {\bibinfo {title} {Relativistic fluid
  dynamics in and out of equilibrium and applications to relativistic nuclear
  collisions}}}\ (\bibinfo  {publisher} {Cambridge University Press},\ \bibinfo
  {address} {Cambridge},\ \bibinfo {year} {2019})\BibitemShut {NoStop}%
\bibitem [{\citenamefont {Bardon}\ \emph {et~al.}(2014)\citenamefont {Bardon},
  \citenamefont {Beattie}, \citenamefont {Luciuk}, \citenamefont {Cairncross},
  \citenamefont {Fine}, \citenamefont {Cheng}, \citenamefont {Edge},
  \citenamefont {Taylor}, \citenamefont {Zhang}, \citenamefont {Trotzky},\ and\
  \citenamefont {Thywissen}}]{bardon2014}%
  \BibitemOpen
  \bibfield  {author} {\bibinfo {author} {\bibfnamefont {A.~B.}\ \bibnamefont
  {Bardon}}, \bibinfo {author} {\bibfnamefont {S.}~\bibnamefont {Beattie}},
  \bibinfo {author} {\bibfnamefont {C.}~\bibnamefont {Luciuk}}, \bibinfo
  {author} {\bibfnamefont {W.}~\bibnamefont {Cairncross}}, \bibinfo {author}
  {\bibfnamefont {D.}~\bibnamefont {Fine}}, \bibinfo {author} {\bibfnamefont
  {N.~S.}\ \bibnamefont {Cheng}}, \bibinfo {author} {\bibfnamefont {G.~J.~A.}\
  \bibnamefont {Edge}}, \bibinfo {author} {\bibfnamefont {E.}~\bibnamefont
  {Taylor}}, \bibinfo {author} {\bibfnamefont {S.}~\bibnamefont {Zhang}},
  \bibinfo {author} {\bibfnamefont {S.}~\bibnamefont {Trotzky}},\ and\ \bibinfo
  {author} {\bibfnamefont {J.~H.}\ \bibnamefont {Thywissen}},\ }\bibfield
  {title} {\bibinfo {title} {{Transverse Demagnetization Dynamics of a Unitary
  Fermi Gas}},\ }\href {https://doi.org/10.1126/science.1247425} {\bibfield
  {journal} {\bibinfo  {journal} {Science}\ }\textbf {\bibinfo {volume}
  {344}},\ \bibinfo {pages} {722} (\bibinfo {year} {2014})}\BibitemShut
  {NoStop}%
\bibitem [{\citenamefont {Luciuk}\ \emph {et~al.}(2017)\citenamefont {Luciuk},
  \citenamefont {Smale}, \citenamefont {B{\"o}ttcher}, \citenamefont {Sharum},
  \citenamefont {Olsen}, \citenamefont {Trotzky}, \citenamefont {Enss},\ and\
  \citenamefont {Thywissen}}]{luciuk2017}%
  \BibitemOpen
  \bibfield  {author} {\bibinfo {author} {\bibfnamefont {C.}~\bibnamefont
  {Luciuk}}, \bibinfo {author} {\bibfnamefont {S.}~\bibnamefont {Smale}},
  \bibinfo {author} {\bibfnamefont {F.}~\bibnamefont {B{\"o}ttcher}}, \bibinfo
  {author} {\bibfnamefont {H.}~\bibnamefont {Sharum}}, \bibinfo {author}
  {\bibfnamefont {B.~A.}\ \bibnamefont {Olsen}}, \bibinfo {author}
  {\bibfnamefont {S.}~\bibnamefont {Trotzky}}, \bibinfo {author} {\bibfnamefont
  {T.}~\bibnamefont {Enss}},\ and\ \bibinfo {author} {\bibfnamefont {J.~H.}\
  \bibnamefont {Thywissen}},\ }\bibfield  {title} {\bibinfo {title}
  {{Observation of quantum-limited spin transport in strongly interacting
  two-dimensional Fermi gases}},\ }\href
  {https://doi.org/10.1103/PhysRevLett.118.130405} {\bibfield  {journal}
  {\bibinfo  {journal} {Phys.\ Rev.\ Lett.}\ }\textbf {\bibinfo {volume}
  {118}},\ \bibinfo {pages} {130405} (\bibinfo {year} {2017})}\BibitemShut
  {NoStop}%
\bibitem [{\citenamefont {Knap}\ \emph {et~al.}(2012)\citenamefont {Knap},
  \citenamefont {Shashi}, \citenamefont {Nishida}, \citenamefont {Imambekov},
  \citenamefont {Abanin},\ and\ \citenamefont {Demler}}]{knap2012}%
  \BibitemOpen
  \bibfield  {author} {\bibinfo {author} {\bibfnamefont {M.}~\bibnamefont
  {Knap}}, \bibinfo {author} {\bibfnamefont {A.}~\bibnamefont {Shashi}},
  \bibinfo {author} {\bibfnamefont {Y.}~\bibnamefont {Nishida}}, \bibinfo
  {author} {\bibfnamefont {A.}~\bibnamefont {Imambekov}}, \bibinfo {author}
  {\bibfnamefont {D.~A.}\ \bibnamefont {Abanin}},\ and\ \bibinfo {author}
  {\bibfnamefont {E.}~\bibnamefont {Demler}},\ }\bibfield  {title} {\bibinfo
  {title} {{Time-dependent impurity in ultracold fermions: Orthogonality
  catastrophe and beyond}},\ }\href {https://doi.org/10.1103/PhysRevX.2.041020}
  {\bibfield  {journal} {\bibinfo  {journal} {Phys.\ Rev.~X}\ }\textbf
  {\bibinfo {volume} {2}},\ \bibinfo {pages} {041020} (\bibinfo {year}
  {2012})}\BibitemShut {NoStop}%
\bibitem [{\citenamefont {Cetina}\ \emph {et~al.}(2016)\citenamefont {Cetina},
  \citenamefont {Jag}, \citenamefont {Lous}, \citenamefont {Fritsche},
  \citenamefont {Walraven}, \citenamefont {Grimm}, \citenamefont {Levinsen},
  \citenamefont {Parish}, \citenamefont {Schmidt}, \citenamefont {Knap},\ and\
  \citenamefont {Demler}}]{cetina2016}%
  \BibitemOpen
  \bibfield  {author} {\bibinfo {author} {\bibfnamefont {M.}~\bibnamefont
  {Cetina}}, \bibinfo {author} {\bibfnamefont {M.}~\bibnamefont {Jag}},
  \bibinfo {author} {\bibfnamefont {R.~S.}\ \bibnamefont {Lous}}, \bibinfo
  {author} {\bibfnamefont {I.}~\bibnamefont {Fritsche}}, \bibinfo {author}
  {\bibfnamefont {J.~T.~M.}\ \bibnamefont {Walraven}}, \bibinfo {author}
  {\bibfnamefont {R.}~\bibnamefont {Grimm}}, \bibinfo {author} {\bibfnamefont
  {J.}~\bibnamefont {Levinsen}}, \bibinfo {author} {\bibfnamefont {M.~M.}\
  \bibnamefont {Parish}}, \bibinfo {author} {\bibfnamefont {R.}~\bibnamefont
  {Schmidt}}, \bibinfo {author} {\bibfnamefont {M.}~\bibnamefont {Knap}},\ and\
  \bibinfo {author} {\bibfnamefont {E.}~\bibnamefont {Demler}},\ }\bibfield
  {title} {\bibinfo {title} {{Ultrafast many-body interferometry of impurities
  coupled to a Fermi sea}},\ }\href {https://doi.org/10.1126/science.aaf5134}
  {\bibfield  {journal} {\bibinfo  {journal} {Science}\ }\textbf {\bibinfo
  {volume} {354}},\ \bibinfo {pages} {96} (\bibinfo {year} {2016})}\BibitemShut
  {NoStop}%
\bibitem [{\citenamefont {Sykes}\ \emph {et~al.}(2014)\citenamefont {Sykes},
  \citenamefont {Corson}, \citenamefont {D'Incao}, \citenamefont {Koller},
  \citenamefont {Greene}, \citenamefont {Rey}, \citenamefont {Hazzard},\ and\
  \citenamefont {Bohn}}]{sykes2014}%
  \BibitemOpen
  \bibfield  {author} {\bibinfo {author} {\bibfnamefont {A.~G.}\ \bibnamefont
  {Sykes}}, \bibinfo {author} {\bibfnamefont {J.~P.}\ \bibnamefont {Corson}},
  \bibinfo {author} {\bibfnamefont {J.~P.}\ \bibnamefont {D'Incao}}, \bibinfo
  {author} {\bibfnamefont {A.~P.}\ \bibnamefont {Koller}}, \bibinfo {author}
  {\bibfnamefont {C.~H.}\ \bibnamefont {Greene}}, \bibinfo {author}
  {\bibfnamefont {A.~M.}\ \bibnamefont {Rey}}, \bibinfo {author} {\bibfnamefont
  {K.~R.~A.}\ \bibnamefont {Hazzard}},\ and\ \bibinfo {author} {\bibfnamefont
  {J.~L.}\ \bibnamefont {Bohn}},\ }\bibfield  {title} {\bibinfo {title}
  {{Quenching to unitarity: Quantum dynamics in a three-dimensional Bose
  gas}},\ }\href {https://doi.org/10.1103/PhysRevA.89.021601} {\bibfield
  {journal} {\bibinfo  {journal} {Phys.\ Rev.~A}\ }\textbf {\bibinfo {volume}
  {89}},\ \bibinfo {pages} {021601(R)} (\bibinfo {year} {2014})}\BibitemShut
  {NoStop}%
\bibitem [{\citenamefont {Fletcher}\ \emph {et~al.}(2017)\citenamefont
  {Fletcher}, \citenamefont {Lopes}, \citenamefont {Man}, \citenamefont
  {Navon}, \citenamefont {Smith}, \citenamefont {Zwierlein},\ and\
  \citenamefont {Hadzibabic}}]{fletcher2017}%
  \BibitemOpen
  \bibfield  {author} {\bibinfo {author} {\bibfnamefont {R.~J.}\ \bibnamefont
  {Fletcher}}, \bibinfo {author} {\bibfnamefont {R.}~\bibnamefont {Lopes}},
  \bibinfo {author} {\bibfnamefont {J.}~\bibnamefont {Man}}, \bibinfo {author}
  {\bibfnamefont {N.}~\bibnamefont {Navon}}, \bibinfo {author} {\bibfnamefont
  {R.~P.}\ \bibnamefont {Smith}}, \bibinfo {author} {\bibfnamefont {M.~W.}\
  \bibnamefont {Zwierlein}},\ and\ \bibinfo {author} {\bibfnamefont
  {Z.}~\bibnamefont {Hadzibabic}},\ }\bibfield  {title} {\bibinfo {title}
  {{Two-and three-body contacts in the unitary Bose gas}},\ }\href
  {https://doi.org/10.1126/science.aai8195} {\bibfield  {journal} {\bibinfo
  {journal} {Science}\ }\textbf {\bibinfo {volume} {355}},\ \bibinfo {pages}
  {377} (\bibinfo {year} {2017})}\BibitemShut {NoStop}%
\bibitem [{\citenamefont {Eigen}\ \emph {et~al.}(2018)\citenamefont {Eigen},
  \citenamefont {Glidden}, \citenamefont {Lopes}, \citenamefont {Cornell},
  \citenamefont {Smith},\ and\ \citenamefont {Hadzibabic}}]{eigen2018}%
  \BibitemOpen
  \bibfield  {author} {\bibinfo {author} {\bibfnamefont {C.}~\bibnamefont
  {Eigen}}, \bibinfo {author} {\bibfnamefont {J.~A.~P.}\ \bibnamefont
  {Glidden}}, \bibinfo {author} {\bibfnamefont {R.}~\bibnamefont {Lopes}},
  \bibinfo {author} {\bibfnamefont {E.~A.}\ \bibnamefont {Cornell}}, \bibinfo
  {author} {\bibfnamefont {R.~P.}\ \bibnamefont {Smith}},\ and\ \bibinfo
  {author} {\bibfnamefont {Z.}~\bibnamefont {Hadzibabic}},\ }\bibfield  {title}
  {\bibinfo {title} {{Universal prethermal dynamics of Bose gases quenched to
  unitarity}},\ }\href {https://doi.org/10.1038/s41586-018-0674-1} {\bibfield
  {journal} {\bibinfo  {journal} {Nature (London)}\ }\textbf {\bibinfo {volume}
  {563}},\ \bibinfo {pages} {221} (\bibinfo {year} {2018})}\BibitemShut
  {NoStop}%
\bibitem [{\citenamefont {Sun}\ \emph {et~al.}(2020)\citenamefont {Sun},
  \citenamefont {Zhang},\ and\ \citenamefont {Zhai}}]{sun2020high}%
  \BibitemOpen
  \bibfield  {author} {\bibinfo {author} {\bibfnamefont {M.}~\bibnamefont
  {Sun}}, \bibinfo {author} {\bibfnamefont {P.}~\bibnamefont {Zhang}},\ and\
  \bibinfo {author} {\bibfnamefont {H.}~\bibnamefont {Zhai}},\ }\bibfield
  {title} {\bibinfo {title} {{High Temperature Virial Expansion to Universal
  Quench Dynamics}},\ }\href {https://doi.org/10.1103/PhysRevLett.125.110404}
  {\bibfield  {journal} {\bibinfo  {journal} {Phys.\ Rev.\ Lett.}\ }\textbf
  {\bibinfo {volume} {125}},\ \bibinfo {pages} {110404} (\bibinfo {year}
  {2020})}\BibitemShut {NoStop}%
\bibitem [{\citenamefont {Drescher}\ \emph {et~al.}(2020)\citenamefont
  {Drescher}, \citenamefont {Salmhofer},\ and\ \citenamefont
  {Enss}}]{drescher2020}%
  \BibitemOpen
  \bibfield  {author} {\bibinfo {author} {\bibfnamefont {M.}~\bibnamefont
  {Drescher}}, \bibinfo {author} {\bibfnamefont {M.}~\bibnamefont
  {Salmhofer}},\ and\ \bibinfo {author} {\bibfnamefont {T.}~\bibnamefont
  {Enss}},\ }\bibfield  {title} {\bibinfo {title} {{Theory of a resonantly
  interacting impurity in a Bose-Einstein condensate}},\ }\href
  {https://doi.org/10.1103/PhysRevResearch.2.032011} {\bibfield  {journal}
  {\bibinfo  {journal} {Phys.\ Rev.\ Res.}\ }\textbf {\bibinfo {volume} {2}},\
  \bibinfo {pages} {032011(R)} (\bibinfo {year} {2020})}\BibitemShut {NoStop}%
\bibitem [{\citenamefont {Drescher}\ \emph {et~al.}(2021)\citenamefont
  {Drescher}, \citenamefont {Salmhofer},\ and\ \citenamefont
  {Enss}}]{drescher2021}%
  \BibitemOpen
  \bibfield  {author} {\bibinfo {author} {\bibfnamefont {M.}~\bibnamefont
  {Drescher}}, \bibinfo {author} {\bibfnamefont {M.}~\bibnamefont
  {Salmhofer}},\ and\ \bibinfo {author} {\bibfnamefont {T.}~\bibnamefont
  {Enss}},\ }\bibfield  {title} {\bibinfo {title} {{Quench Dynamics of the
  Ideal Bose Polaron at Zero and Nonzero Temperatures}},\ }\href
  {https://doi.org/10.1103/PhysRevA.103.033317} {\bibfield  {journal} {\bibinfo
   {journal} {Phys.\ Rev.~A}\ }\textbf {\bibinfo {volume} {103}},\ \bibinfo
  {pages} {033317} (\bibinfo {year} {2021})}\BibitemShut {NoStop}%
\bibitem [{\citenamefont {Skou}\ \emph {et~al.}(2021)\citenamefont {Skou},
  \citenamefont {Skov}, \citenamefont {J{\o}rgensen}, \citenamefont {Nielsen},
  \citenamefont {Camacho-Guardian}, \citenamefont {Pohl}, \citenamefont
  {Bruun},\ and\ \citenamefont {Arlt}}]{skou2021}%
  \BibitemOpen
  \bibfield  {author} {\bibinfo {author} {\bibfnamefont {M.~G.}\ \bibnamefont
  {Skou}}, \bibinfo {author} {\bibfnamefont {T.~G.}\ \bibnamefont {Skov}},
  \bibinfo {author} {\bibfnamefont {N.~B.}\ \bibnamefont {J{\o}rgensen}},
  \bibinfo {author} {\bibfnamefont {K.~K.}\ \bibnamefont {Nielsen}}, \bibinfo
  {author} {\bibfnamefont {A.}~\bibnamefont {Camacho-Guardian}}, \bibinfo
  {author} {\bibfnamefont {T.}~\bibnamefont {Pohl}}, \bibinfo {author}
  {\bibfnamefont {G.~M.}\ \bibnamefont {Bruun}},\ and\ \bibinfo {author}
  {\bibfnamefont {J.~J.}\ \bibnamefont {Arlt}},\ }\bibfield  {title} {\bibinfo
  {title} {{Non-equilibrium quantum dynamics and formation of the Bose
  polaron}},\ }\href {https://doi.org/10.1038/s41567-021-01184-5} {\bibfield
  {journal} {\bibinfo  {journal} {Nat. Phys.}\ }\textbf {\bibinfo {volume}
  {17}},\ \bibinfo {pages} {731} (\bibinfo {year} {2021})}\BibitemShut
  {NoStop}%
\bibitem [{\citenamefont {Qi}\ \emph {et~al.}(2021)\citenamefont {Qi},
  \citenamefont {Shi},\ and\ \citenamefont {Zhai}}]{qi2021maximum}%
  \BibitemOpen
  \bibfield  {author} {\bibinfo {author} {\bibfnamefont {R.}~\bibnamefont
  {Qi}}, \bibinfo {author} {\bibfnamefont {Z.}~\bibnamefont {Shi}},\ and\
  \bibinfo {author} {\bibfnamefont {H.}~\bibnamefont {Zhai}},\ }\bibfield
  {title} {\bibinfo {title} {{Maximum Energy Growth Rate in Dilute Quantum
  Gases}},\ }\href {https://doi.org/10.1103/PhysRevLett.126.240401} {\bibfield
  {journal} {\bibinfo  {journal} {Phys.\ Rev.\ Lett.}\ }\textbf {\bibinfo
  {volume} {126}},\ \bibinfo {pages} {240401} (\bibinfo {year}
  {2021})}\BibitemShut {NoStop}%
\bibitem [{\citenamefont {Maki}\ \emph {et~al.}(2022)\citenamefont {Maki},
  \citenamefont {Zhang},\ and\ \citenamefont {Zhou}}]{maki2022dynamics}%
  \BibitemOpen
  \bibfield  {author} {\bibinfo {author} {\bibfnamefont {J.}~\bibnamefont
  {Maki}}, \bibinfo {author} {\bibfnamefont {S.}~\bibnamefont {Zhang}},\ and\
  \bibinfo {author} {\bibfnamefont {F.}~\bibnamefont {Zhou}},\ }\bibfield
  {title} {\bibinfo {title} {{Dynamics of strongly interacting Fermi gases with
  time-dependent interactions: Consequence of conformal symmetry}},\ }\href
  {https://doi.org/10.1103/PhysRevLett.128.040401} {\bibfield  {journal}
  {\bibinfo  {journal} {Phys.\ Rev.\ Lett.}\ }\textbf {\bibinfo {volume}
  {128}},\ \bibinfo {pages} {040401} (\bibinfo {year} {2022})}\BibitemShut
  {NoStop}%
\bibitem [{\citenamefont {Tan}(2008{\natexlab{a}})}]{tan2008energetics}%
  \BibitemOpen
  \bibfield  {author} {\bibinfo {author} {\bibfnamefont {S.}~\bibnamefont
  {Tan}},\ }\bibfield  {title} {\bibinfo {title} {{Energetics of a strongly
  correlated Fermi gas}},\ }\href {https://doi.org/10.1016/j.aop.2008.03.004}
  {\bibfield  {journal} {\bibinfo  {journal} {Ann.\ Phys.\ (N.Y.)}\ }\textbf
  {\bibinfo {volume} {323}},\ \bibinfo {pages} {2952} (\bibinfo {year}
  {2008}{\natexlab{a}})}\BibitemShut {NoStop}%
\bibitem [{\citenamefont {Tan}(2008{\natexlab{b}})}]{tan2008large}%
  \BibitemOpen
  \bibfield  {author} {\bibinfo {author} {\bibfnamefont {S.}~\bibnamefont
  {Tan}},\ }\bibfield  {title} {\bibinfo {title} {{Large momentum part of a
  strongly correlated Fermi gas}},\ }\href
  {https://doi.org/10.1016/j.aop.2008.03.005} {\bibfield  {journal} {\bibinfo
  {journal} {Ann.\ Phys.\ (N.Y.)}\ }\textbf {\bibinfo {volume} {323}},\
  \bibinfo {pages} {2971} (\bibinfo {year} {2008}{\natexlab{b}})}\BibitemShut
  {NoStop}%
\bibitem [{\citenamefont {Pitaevskii}\ and\ \citenamefont
  {Rosch}(1997)}]{pitaevskii1997}%
  \BibitemOpen
  \bibfield  {author} {\bibinfo {author} {\bibfnamefont {L.~P.}\ \bibnamefont
  {Pitaevskii}}\ and\ \bibinfo {author} {\bibfnamefont {A.}~\bibnamefont
  {Rosch}},\ }\bibfield  {title} {\bibinfo {title} {Breathing modes and hidden
  symmetry of trapped atoms in two dimensions},\ }\href
  {https://doi.org/10.1103/PhysRevA.55.R853} {\bibfield  {journal} {\bibinfo
  {journal} {Phys.\ Rev.~A}\ }\textbf {\bibinfo {volume} {55}},\ \bibinfo
  {pages} {R853} (\bibinfo {year} {1997})}\BibitemShut {NoStop}%
\bibitem [{\citenamefont {Werner}\ and\ \citenamefont
  {Castin}(2006)}]{werner2006unitary}%
  \BibitemOpen
  \bibfield  {author} {\bibinfo {author} {\bibfnamefont {F.}~\bibnamefont
  {Werner}}\ and\ \bibinfo {author} {\bibfnamefont {Y.}~\bibnamefont
  {Castin}},\ }\bibfield  {title} {\bibinfo {title} {{Unitary gas in an
  isotropic harmonic trap: Symmetry properties and applications}},\ }\href
  {https://doi.org/10.1103/PhysRevA.74.053604} {\bibfield  {journal} {\bibinfo
  {journal} {Phys.\ Rev.~A}\ }\textbf {\bibinfo {volume} {74}},\ \bibinfo
  {pages} {053604} (\bibinfo {year} {2006})}\BibitemShut {NoStop}%
\bibitem [{\citenamefont {Gritsev}\ \emph {et~al.}(2010)\citenamefont
  {Gritsev}, \citenamefont {Barmettler},\ and\ \citenamefont
  {Demler}}]{gritsev2010}%
  \BibitemOpen
  \bibfield  {author} {\bibinfo {author} {\bibfnamefont {V.}~\bibnamefont
  {Gritsev}}, \bibinfo {author} {\bibfnamefont {P.}~\bibnamefont
  {Barmettler}},\ and\ \bibinfo {author} {\bibfnamefont {E.}~\bibnamefont
  {Demler}},\ }\bibfield  {title} {\bibinfo {title} {Scaling approach to
  quantum non-equilibrium dynamics of many-body systems},\ }\href
  {https://doi.org/10.1088/1367-2630/12/11/113005} {\bibfield  {journal}
  {\bibinfo  {journal} {New J. Phys.}\ }\textbf {\bibinfo {volume} {12}},\
  \bibinfo {pages} {113005} (\bibinfo {year} {2010})}\BibitemShut {NoStop}%
\bibitem [{\citenamefont {Balslev}\ and\ \citenamefont
  {Combes}(1971)}]{balslev1971spectral}%
  \BibitemOpen
  \bibfield  {author} {\bibinfo {author} {\bibfnamefont {E.}~\bibnamefont
  {Balslev}}\ and\ \bibinfo {author} {\bibfnamefont {J.~M.}\ \bibnamefont
  {Combes}},\ }\bibfield  {title} {\bibinfo {title} {{Spectral properties of
  Schr{\"o}dinger Hamiltonians with dilation analytic potentials}},\ }\href
  {https://doi.org/10.1007/BF01877511} {\bibfield  {journal} {\bibinfo
  {journal} {Commun. Math. Phys.}\ }\textbf {\bibinfo {volume} {22}},\ \bibinfo
  {pages} {280} (\bibinfo {year} {1971})}\BibitemShut {NoStop}%
\bibitem [{\citenamefont {Reed}\ and\ \citenamefont
  {Simon}(1978)}]{reed1978vol4}%
  \BibitemOpen
  \bibfield  {author} {\bibinfo {author} {\bibfnamefont {M.}~\bibnamefont
  {Reed}}\ and\ \bibinfo {author} {\bibfnamefont {B.}~\bibnamefont {Simon}},\
  }\href@noop {} {\emph {\bibinfo {title} {{Methods of Modern Mathematical
  Physics: Analysis of Operators}}}},\ Vol.~\bibinfo {volume} {4}\ (\bibinfo
  {publisher} {Academic Press},\ \bibinfo {address} {San Diego},\ \bibinfo
  {year} {1978})\BibitemShut {NoStop}%
\bibitem [{\citenamefont {Bach}\ \emph {et~al.}(1998)\citenamefont {Bach},
  \citenamefont {Fr{\"o}hlich},\ and\ \citenamefont {Sigal}}]{bach1998quantum}%
  \BibitemOpen
  \bibfield  {author} {\bibinfo {author} {\bibfnamefont {V.}~\bibnamefont
  {Bach}}, \bibinfo {author} {\bibfnamefont {J.}~\bibnamefont {Fr{\"o}hlich}},\
  and\ \bibinfo {author} {\bibfnamefont {I.~M.}\ \bibnamefont {Sigal}},\
  }\bibfield  {title} {\bibinfo {title} {Quantum electrodynamics of confined
  nonrelativistic particles},\ }\href {https://doi.org/10.1006/aima.1998.1734}
  {\bibfield  {journal} {\bibinfo  {journal} {Adv. in Math.}\ }\textbf
  {\bibinfo {volume} {137}},\ \bibinfo {pages} {299} (\bibinfo {year}
  {1998})}\BibitemShut {NoStop}%
\bibitem [{\citenamefont {Busch}\ \emph {et~al.}(1998)\citenamefont {Busch},
  \citenamefont {Englert}, \citenamefont {Rza{\.z}ewski},\ and\ \citenamefont
  {Wilkens}}]{busch1998}%
  \BibitemOpen
  \bibfield  {author} {\bibinfo {author} {\bibfnamefont {T.}~\bibnamefont
  {Busch}}, \bibinfo {author} {\bibfnamefont {B.-G.}\ \bibnamefont {Englert}},
  \bibinfo {author} {\bibfnamefont {K.}~\bibnamefont {Rza{\.z}ewski}},\ and\
  \bibinfo {author} {\bibfnamefont {M.}~\bibnamefont {Wilkens}},\ }\bibfield
  {title} {\bibinfo {title} {Two cold atoms in a harmonic trap},\ }\href
  {https://doi.org/10.1023/A:1018705520999} {\bibfield  {journal} {\bibinfo
  {journal} {Found. Phys.}\ }\textbf {\bibinfo {volume} {28}},\ \bibinfo
  {pages} {549} (\bibinfo {year} {1998})}\BibitemShut {NoStop}%
\bibitem [{\citenamefont {Kerin}\ and\ \citenamefont
  {Martin}(2020)}]{kerin2020two}%
  \BibitemOpen
  \bibfield  {author} {\bibinfo {author} {\bibfnamefont {A.~D.}\ \bibnamefont
  {Kerin}}\ and\ \bibinfo {author} {\bibfnamefont {A.~M.}\ \bibnamefont
  {Martin}},\ }\bibfield  {title} {\bibinfo {title} {Two-body quench dynamics
  of harmonically trapped interacting particles},\ }\href
  {https://doi.org/10.1103/PhysRevA.102.023311} {\bibfield  {journal} {\bibinfo
   {journal} {Phys.\ Rev.~A}\ }\textbf {\bibinfo {volume} {102}},\ \bibinfo
  {pages} {023311} (\bibinfo {year} {2020})}\BibitemShut {NoStop}%
\bibitem [{\citenamefont {Kehrberger}\ \emph {et~al.}(2018)\citenamefont
  {Kehrberger}, \citenamefont {Bolsinger},\ and\ \citenamefont
  {Schmelcher}}]{kehrberger2018}%
  \BibitemOpen
  \bibfield  {author} {\bibinfo {author} {\bibfnamefont {L.~M.~A.}\
  \bibnamefont {Kehrberger}}, \bibinfo {author} {\bibfnamefont {V.~J.}\
  \bibnamefont {Bolsinger}},\ and\ \bibinfo {author} {\bibfnamefont
  {P.}~\bibnamefont {Schmelcher}},\ }\bibfield  {title} {\bibinfo {title}
  {Quantum dynamics of two trapped bosons following infinite interaction
  quenches},\ }\href {https://doi.org/10.1103/PhysRevA.97.013606} {\bibfield
  {journal} {\bibinfo  {journal} {Physical Review A}\ }\textbf {\bibinfo
  {volume} {97}},\ \bibinfo {pages} {013606} (\bibinfo {year}
  {2018})}\BibitemShut {NoStop}%
\bibitem [{\citenamefont {Nishida}\ and\ \citenamefont
  {Son}(2007)}]{nishida2007nonrel}%
  \BibitemOpen
  \bibfield  {author} {\bibinfo {author} {\bibfnamefont {Y.}~\bibnamefont
  {Nishida}}\ and\ \bibinfo {author} {\bibfnamefont {D.~T.}\ \bibnamefont
  {Son}},\ }\bibfield  {title} {\bibinfo {title} {{Nonrelativistic conformal
  field theories}},\ }\href {https://doi.org/10.1103/PhysRevD.76.086004}
  {\bibfield  {journal} {\bibinfo  {journal} {Phys.\ Rev.~D}\ }\textbf
  {\bibinfo {volume} {76}},\ \bibinfo {pages} {086004} (\bibinfo {year}
  {2007})}\BibitemShut {NoStop}%
\bibitem [{\citenamefont {Werner}\ and\ \citenamefont
  {Castin}(2012)}]{werner2012}%
  \BibitemOpen
  \bibfield  {author} {\bibinfo {author} {\bibfnamefont {F.}~\bibnamefont
  {Werner}}\ and\ \bibinfo {author} {\bibfnamefont {Y.}~\bibnamefont
  {Castin}},\ }\bibfield  {title} {\bibinfo {title} {{General relations for
  quantum gases in two and three dimensions. Two-component fermions}},\ }\href
  {https://doi.org/10.1103/PhysRevA.86.013626} {\bibfield  {journal} {\bibinfo
  {journal} {Phys.\ Rev.~A}\ }\textbf {\bibinfo {volume} {86}},\ \bibinfo
  {pages} {013626} (\bibinfo {year} {2012})}\BibitemShut {NoStop}%
\bibitem [{\citenamefont {Moroz}(2012)}]{moroz2012scale}%
  \BibitemOpen
  \bibfield  {author} {\bibinfo {author} {\bibfnamefont {S.}~\bibnamefont
  {Moroz}},\ }\bibfield  {title} {\bibinfo {title} {{Scale-invariant Fermi gas
  in a time-dependent harmonic potential}},\ }\href
  {https://doi.org/10.1103/PhysRevA.86.011601} {\bibfield  {journal} {\bibinfo
  {journal} {Phys.\ Rev.~A}\ }\textbf {\bibinfo {volume} {86}},\ \bibinfo
  {pages} {011601(R)} (\bibinfo {year} {2012})}\BibitemShut {NoStop}%
\bibitem [{\citenamefont {Parish}\ and\ \citenamefont
  {Levinsen}(2016)}]{parish2016quantum}%
  \BibitemOpen
  \bibfield  {author} {\bibinfo {author} {\bibfnamefont {M.~M.}\ \bibnamefont
  {Parish}}\ and\ \bibinfo {author} {\bibfnamefont {J.}~\bibnamefont
  {Levinsen}},\ }\bibfield  {title} {\bibinfo {title} {{Quantum dynamics of
  impurities coupled to a Fermi sea}},\ }\href
  {https://doi.org/10.1103/PhysRevB.94.184303} {\bibfield  {journal} {\bibinfo
  {journal} {Phys.\ Rev.~B}\ }\textbf {\bibinfo {volume} {94}},\ \bibinfo
  {pages} {184303} (\bibinfo {year} {2016})}\BibitemShut {NoStop}%
\bibitem [{\citenamefont {Sch{\"a}fer}\ and\ \citenamefont
  {Chafin}(2012)}]{schaefer2012scaling}%
  \BibitemOpen
  \bibfield  {author} {\bibinfo {author} {\bibfnamefont {T.}~\bibnamefont
  {Sch{\"a}fer}}\ and\ \bibinfo {author} {\bibfnamefont {C.}~\bibnamefont
  {Chafin}},\ }\bibfield  {title} {\bibinfo {title} {{Scaling Flows and
  Dissipation in the Dilute Fermi Gas at Unitarity}},\ }in\ \href@noop {}
  {\emph {\bibinfo {booktitle} {{The BCS-BEC Crossover and the Unitary Fermi
  Gas}}}},\ \bibinfo {editor} {edited by\ \bibinfo {editor} {\bibfnamefont
  {W.}~\bibnamefont {Zwerger}}}\ (\bibinfo  {publisher} {Springer},\ \bibinfo
  {address} {Berlin},\ \bibinfo {year} {2012})\ Chap.~\bibinfo {chapter} {10},
  p.\ \bibinfo {pages} {375}\BibitemShut {NoStop}%
\bibitem [{\citenamefont {Berges}\ \emph {et~al.}(2008)\citenamefont {Berges},
  \citenamefont {Rothkopf},\ and\ \citenamefont
  {Schmidt}}]{berges2008nonthermal}%
  \BibitemOpen
  \bibfield  {author} {\bibinfo {author} {\bibfnamefont {J.}~\bibnamefont
  {Berges}}, \bibinfo {author} {\bibfnamefont {A.}~\bibnamefont {Rothkopf}},\
  and\ \bibinfo {author} {\bibfnamefont {J.}~\bibnamefont {Schmidt}},\
  }\bibfield  {title} {\bibinfo {title} {Nonthermal fixed points: effective
  weak coupling for strongly correlated systems far from equilibrium},\ }\href
  {https://doi.org/10.1103/PhysRevLett.101.041603} {\bibfield  {journal}
  {\bibinfo  {journal} {Phys.\ Rev.\ Lett.}\ }\textbf {\bibinfo {volume}
  {101}},\ \bibinfo {pages} {041603} (\bibinfo {year} {2008})}\BibitemShut
  {NoStop}%
\bibitem [{\citenamefont {Murthy}\ \emph {et~al.}(2019)\citenamefont {Murthy},
  \citenamefont {Defenu}, \citenamefont {Bayha}, \citenamefont {Holten},
  \citenamefont {Preiss}, \citenamefont {Enss},\ and\ \citenamefont
  {Jochim}}]{murthy2019}%
  \BibitemOpen
  \bibfield  {author} {\bibinfo {author} {\bibfnamefont {P.~A.}\ \bibnamefont
  {Murthy}}, \bibinfo {author} {\bibfnamefont {N.}~\bibnamefont {Defenu}},
  \bibinfo {author} {\bibfnamefont {L.}~\bibnamefont {Bayha}}, \bibinfo
  {author} {\bibfnamefont {M.}~\bibnamefont {Holten}}, \bibinfo {author}
  {\bibfnamefont {P.~M.}\ \bibnamefont {Preiss}}, \bibinfo {author}
  {\bibfnamefont {T.}~\bibnamefont {Enss}},\ and\ \bibinfo {author}
  {\bibfnamefont {S.}~\bibnamefont {Jochim}},\ }\bibfield  {title} {\bibinfo
  {title} {{Quantum scale anomaly and spatial coherence in a 2D Fermi
  superfluid}},\ }\href {https://doi.org/10.1126/science.aau4402} {\bibfield
  {journal} {\bibinfo  {journal} {Science}\ }\textbf {\bibinfo {volume}
  {365}},\ \bibinfo {pages} {268} (\bibinfo {year} {2019})}\BibitemShut
  {NoStop}%
\bibitem [{\citenamefont {Olshanii}\ \emph {et~al.}(2010)\citenamefont
  {Olshanii}, \citenamefont {Perrin},\ and\ \citenamefont
  {Lorent}}]{olshanii2010}%
  \BibitemOpen
  \bibfield  {author} {\bibinfo {author} {\bibfnamefont {M.}~\bibnamefont
  {Olshanii}}, \bibinfo {author} {\bibfnamefont {H.}~\bibnamefont {Perrin}},\
  and\ \bibinfo {author} {\bibfnamefont {V.}~\bibnamefont {Lorent}},\
  }\bibfield  {title} {\bibinfo {title} {Example of a quantum anomaly in the
  physics of ultracold gases},\ }\href
  {https://doi.org/10.1103/PhysRevLett.105.095302} {\bibfield  {journal}
  {\bibinfo  {journal} {Phys.\ Rev.\ Lett.}\ }\textbf {\bibinfo {volume}
  {105}},\ \bibinfo {pages} {095302} (\bibinfo {year} {2010})}\BibitemShut
  {NoStop}%
\bibitem [{\citenamefont {Hofmann}(2012)}]{hofmann2012}%
  \BibitemOpen
  \bibfield  {author} {\bibinfo {author} {\bibfnamefont {J.}~\bibnamefont
  {Hofmann}},\ }\bibfield  {title} {\bibinfo {title} {{Quantum Anomaly,
  Universal Relations, and Breathing Mode of a Two-Dimensional Fermi Gas}},\
  }\href {https://doi.org/10.1103/PhysRevLett.108.185303} {\bibfield  {journal}
  {\bibinfo  {journal} {Phys.\ Rev.\ Lett.}\ }\textbf {\bibinfo {volume}
  {108}},\ \bibinfo {pages} {185303} (\bibinfo {year} {2012})}\BibitemShut
  {NoStop}%
\bibitem [{\citenamefont {Holten}\ \emph {et~al.}(2018)\citenamefont {Holten},
  \citenamefont {Bayha}, \citenamefont {Klein}, \citenamefont {Murthy},
  \citenamefont {Preiss},\ and\ \citenamefont {Jochim}}]{holten2018}%
  \BibitemOpen
  \bibfield  {author} {\bibinfo {author} {\bibfnamefont {M.}~\bibnamefont
  {Holten}}, \bibinfo {author} {\bibfnamefont {L.}~\bibnamefont {Bayha}},
  \bibinfo {author} {\bibfnamefont {A.~C.}\ \bibnamefont {Klein}}, \bibinfo
  {author} {\bibfnamefont {P.~A.}\ \bibnamefont {Murthy}}, \bibinfo {author}
  {\bibfnamefont {P.~M.}\ \bibnamefont {Preiss}},\ and\ \bibinfo {author}
  {\bibfnamefont {S.}~\bibnamefont {Jochim}},\ }\bibfield  {title} {\bibinfo
  {title} {{Anomalous breaking of scale invariance in a two-dimensional Fermi
  gas}},\ }\href {https://doi.org/10.1103/PhysRevLett.121.120401} {\bibfield
  {journal} {\bibinfo  {journal} {Phys.\ Rev.\ Lett.}\ }\textbf {\bibinfo
  {volume} {121}},\ \bibinfo {pages} {120401} (\bibinfo {year}
  {2018})}\BibitemShut {NoStop}%
\bibitem [{\citenamefont {Peppler}\ \emph {et~al.}(2018)\citenamefont
  {Peppler}, \citenamefont {Dyke}, \citenamefont {Zamorano}, \citenamefont
  {Herrera}, \citenamefont {Hoinka},\ and\ \citenamefont {Vale}}]{peppler2018}%
  \BibitemOpen
  \bibfield  {author} {\bibinfo {author} {\bibfnamefont {T.}~\bibnamefont
  {Peppler}}, \bibinfo {author} {\bibfnamefont {P.}~\bibnamefont {Dyke}},
  \bibinfo {author} {\bibfnamefont {M.}~\bibnamefont {Zamorano}}, \bibinfo
  {author} {\bibfnamefont {I.}~\bibnamefont {Herrera}}, \bibinfo {author}
  {\bibfnamefont {S.}~\bibnamefont {Hoinka}},\ and\ \bibinfo {author}
  {\bibfnamefont {C.~J.}\ \bibnamefont {Vale}},\ }\bibfield  {title} {\bibinfo
  {title} {{Quantum anomaly and 2D-3D crossover in strongly interacting Fermi
  gases}},\ }\href {https://doi.org/10.1103/PhysRevLett.121.120402} {\bibfield
  {journal} {\bibinfo  {journal} {Phys.\ Rev.\ Lett.}\ }\textbf {\bibinfo
  {volume} {121}},\ \bibinfo {pages} {120402} (\bibinfo {year}
  {2018})}\BibitemShut {NoStop}%
\bibitem [{\citenamefont {Syed}\ \emph {et~al.}(2021)\citenamefont {Syed},
  \citenamefont {Enss},\ and\ \citenamefont {Defenu}}]{syed2021dynamical}%
  \BibitemOpen
  \bibfield  {author} {\bibinfo {author} {\bibfnamefont {M.}~\bibnamefont
  {Syed}}, \bibinfo {author} {\bibfnamefont {T.}~\bibnamefont {Enss}},\ and\
  \bibinfo {author} {\bibfnamefont {N.}~\bibnamefont {Defenu}},\ }\bibfield
  {title} {\bibinfo {title} {Dynamical quantum phase transition in a bosonic
  system with long-range interactions},\ }\href
  {https://doi.org/10.1103/PhysRevB.103.064306} {\bibfield  {journal} {\bibinfo
   {journal} {Phys.\ Rev.~B}\ }\textbf {\bibinfo {volume} {103}},\ \bibinfo
  {pages} {064306} (\bibinfo {year} {2021})}\BibitemShut {NoStop}%
\bibitem [{\citenamefont {Blume}(2012)}]{blume2012few}%
  \BibitemOpen
  \bibfield  {author} {\bibinfo {author} {\bibfnamefont {D.}~\bibnamefont
  {Blume}},\ }\bibfield  {title} {\bibinfo {title} {Few-body physics with
  ultracold atomic and molecular systems in traps},\ }\href
  {https://doi.org/10.1088/0034-4885/75/4/046401} {\bibfield  {journal}
  {\bibinfo  {journal} {Rep.\ Prog.\ Phys.}\ }\textbf {\bibinfo {volume}
  {75}},\ \bibinfo {pages} {046401} (\bibinfo {year} {2012})}\BibitemShut
  {NoStop}%
\bibitem [{\citenamefont {Nishida}\ \emph {et~al.}(2008)\citenamefont
  {Nishida}, \citenamefont {Son},\ and\ \citenamefont
  {Tan}}]{nishida2008universal}%
  \BibitemOpen
  \bibfield  {author} {\bibinfo {author} {\bibfnamefont {Y.}~\bibnamefont
  {Nishida}}, \bibinfo {author} {\bibfnamefont {D.~T.}\ \bibnamefont {Son}},\
  and\ \bibinfo {author} {\bibfnamefont {S.}~\bibnamefont {Tan}},\ }\bibfield
  {title} {\bibinfo {title} {{Universal Fermi gas with two-and three-body
  resonances}},\ }\href {https://doi.org/10.1103/PhysRevLett.100.090405}
  {\bibfield  {journal} {\bibinfo  {journal} {Phys.\ Rev.\ Lett.}\ }\textbf
  {\bibinfo {volume} {100}},\ \bibinfo {pages} {090405} (\bibinfo {year}
  {2008})}\BibitemShut {NoStop}%
\bibitem [{\citenamefont
  {Pricoupenko}(2006{\natexlab{a}})}]{pricoupenko2006modeling}%
  \BibitemOpen
  \bibfield  {author} {\bibinfo {author} {\bibfnamefont {L.}~\bibnamefont
  {Pricoupenko}},\ }\bibfield  {title} {\bibinfo {title} {{Modeling
  interactions for resonant $p$-wave scattering}},\ }\href
  {https://doi.org/10.1103/PhysRevLett.96.050401} {\bibfield  {journal}
  {\bibinfo  {journal} {Phys.\ Rev.\ Lett.}\ }\textbf {\bibinfo {volume}
  {96}},\ \bibinfo {pages} {050401} (\bibinfo {year}
  {2006}{\natexlab{a}})}\BibitemShut {NoStop}%
\bibitem [{\citenamefont
  {Pricoupenko}(2006{\natexlab{b}})}]{pricoupenko2006pseudopotential}%
  \BibitemOpen
  \bibfield  {author} {\bibinfo {author} {\bibfnamefont {L.}~\bibnamefont
  {Pricoupenko}},\ }\bibfield  {title} {\bibinfo {title} {Pseudopotential in
  resonant regimes},\ }\href {https://doi.org/10.1103/PhysRevA.73.012701}
  {\bibfield  {journal} {\bibinfo  {journal} {Phys.\ Rev.~A}\ }\textbf
  {\bibinfo {volume} {73}},\ \bibinfo {pages} {012701} (\bibinfo {year}
  {2006}{\natexlab{b}})}\BibitemShut {NoStop}%
\bibitem [{\citenamefont {Zhukov}\ \emph {et~al.}(1993)\citenamefont {Zhukov},
  \citenamefont {Danilin}, \citenamefont {Fedorov}, \citenamefont {Bang},
  \citenamefont {Thompson},\ and\ \citenamefont {Vaagen}}]{zhukov1993}%
  \BibitemOpen
  \bibfield  {author} {\bibinfo {author} {\bibfnamefont {M.~V.}\ \bibnamefont
  {Zhukov}}, \bibinfo {author} {\bibfnamefont {B.~V.}\ \bibnamefont {Danilin}},
  \bibinfo {author} {\bibfnamefont {D.}~\bibnamefont {Fedorov}}, \bibinfo
  {author} {\bibfnamefont {J.~M.}\ \bibnamefont {Bang}}, \bibinfo {author}
  {\bibfnamefont {I.~J.}\ \bibnamefont {Thompson}},\ and\ \bibinfo {author}
  {\bibfnamefont {J.~S.}\ \bibnamefont {Vaagen}},\ }\bibfield  {title}
  {\bibinfo {title} {{Bound state properties of Borromean halo nuclei: $^6$He
  and $^{11}$Li}},\ }\href {https://doi.org/10.1016/0370-1573(93)90141-Y}
  {\bibfield  {journal} {\bibinfo  {journal} {Phys.\ Rep.}\ }\textbf {\bibinfo
  {volume} {231}},\ \bibinfo {pages} {151} (\bibinfo {year}
  {1993})}\BibitemShut {NoStop}%
\bibitem [{\citenamefont {Castin}\ and\ \citenamefont
  {Werner}(2012)}]{castin2012unitary}%
  \BibitemOpen
  \bibfield  {author} {\bibinfo {author} {\bibfnamefont {Y.}~\bibnamefont
  {Castin}}\ and\ \bibinfo {author} {\bibfnamefont {F.}~\bibnamefont
  {Werner}},\ }\bibfield  {title} {\bibinfo {title} {{The Unitary Gas and its
  Symmetry Properties}},\ }in\ \href@noop {} {\emph {\bibinfo {booktitle} {{The
  BCS-BEC Crossover and the Unitary Fermi Gas}}}},\ \bibinfo {editor} {edited
  by\ \bibinfo {editor} {\bibfnamefont {W.}~\bibnamefont {Zwerger}}}\ (\bibinfo
   {publisher} {Springer},\ \bibinfo {address} {Berlin},\ \bibinfo {year}
  {2012})\ Chap.~\bibinfo {chapter} {5}, p.\ \bibinfo {pages} {127}\BibitemShut
  {NoStop}%
\bibitem [{\citenamefont {Enss}\ and\ \citenamefont
  {Haussmann}(2012)}]{enss2012spin}%
  \BibitemOpen
  \bibfield  {author} {\bibinfo {author} {\bibfnamefont {T.}~\bibnamefont
  {Enss}}\ and\ \bibinfo {author} {\bibfnamefont {R.}~\bibnamefont
  {Haussmann}},\ }\bibfield  {title} {\bibinfo {title} {{Quantum Mechanical
  Limitations to Spin Transport in the Unitary Fermi Gas}},\ }\href
  {https://doi.org/10.1103/PhysRevLett.109.195303} {\bibfield  {journal}
  {\bibinfo  {journal} {Phys.\ Rev.\ Lett.}\ }\textbf {\bibinfo {volume}
  {109}},\ \bibinfo {pages} {195303} (\bibinfo {year} {2012})}\BibitemShut
  {NoStop}%
\bibitem [{\citenamefont {Hongo}\ and\ \citenamefont
  {Son}(2022)}]{hongo2022universal}%
  \BibitemOpen
  \bibfield  {author} {\bibinfo {author} {\bibfnamefont {M.}~\bibnamefont
  {Hongo}}\ and\ \bibinfo {author} {\bibfnamefont {D.~T.}\ \bibnamefont
  {Son}},\ }\bibfield  {title} {\bibinfo {title} {{Universal Properties of
  Weakly Bound Two-Neutron Halo Nuclei}},\ }\href
  {https://doi.org/10.1103/PhysRevLett.128.212501} {\bibfield  {journal}
  {\bibinfo  {journal} {Phys.\ Rev.\ Lett.}\ }\textbf {\bibinfo {volume}
  {128}},\ \bibinfo {pages} {212501} (\bibinfo {year} {2022})}\BibitemShut
  {NoStop}%
\bibitem [{\citenamefont {Zwerger}(2019)}]{zwerger2019quantum}%
  \BibitemOpen
  \bibfield  {author} {\bibinfo {author} {\bibfnamefont {W.}~\bibnamefont
  {Zwerger}},\ }\bibfield  {title} {\bibinfo {title} {Quantum-unbinding near a
  zero temperature liquid-gas transition},\ }\href
  {https://doi.org/10.1088/1742-5468/ab3ccc} {\bibfield  {journal} {\bibinfo
  {journal} {Journal of Statistical Mechanics: Theory and Experiment}\ }\textbf
  {\bibinfo {volume} {2019}},\ \bibinfo {pages} {103104} (\bibinfo {year}
  {2019})}\BibitemShut {NoStop}%
\bibitem [{\citenamefont {Georgi}(2007)}]{georgi2007}%
  \BibitemOpen
  \bibfield  {author} {\bibinfo {author} {\bibfnamefont {H.}~\bibnamefont
  {Georgi}},\ }\bibfield  {title} {\bibinfo {title} {Unparticle physics},\
  }\href {https://doi.org/10.1103/PhysRevLett.98.221601} {\bibfield  {journal}
  {\bibinfo  {journal} {Phys.\ Rev.\ Lett.}\ }\textbf {\bibinfo {volume}
  {98}},\ \bibinfo {pages} {221601} (\bibinfo {year} {2007})}\BibitemShut
  {NoStop}%
\bibitem [{\citenamefont {Hammer}\ and\ \citenamefont
  {Son}(2021)}]{hammer2021unnuclear}%
  \BibitemOpen
  \bibfield  {author} {\bibinfo {author} {\bibfnamefont {H.-W.}\ \bibnamefont
  {Hammer}}\ and\ \bibinfo {author} {\bibfnamefont {D.~T.}\ \bibnamefont
  {Son}},\ }\bibfield  {title} {\bibinfo {title} {{Unnuclear physics: Conformal
  symmetry in nuclear reactions}},\ }\href
  {https://doi.org/10.1073/pnas.2108716118} {\bibfield  {journal} {\bibinfo
  {journal} {Proc. Nat. Acad. Sci. USA}\ }\textbf {\bibinfo {volume} {118}},\
  \bibinfo {pages} {e2108716118} (\bibinfo {year} {2021})}\BibitemShut
  {NoStop}%
\bibitem [{\citenamefont {Colussi}\ \emph {et~al.}(2018)\citenamefont
  {Colussi}, \citenamefont {Corson},\ and\ \citenamefont
  {D’Incao}}]{colussi2018dynamics}%
  \BibitemOpen
  \bibfield  {author} {\bibinfo {author} {\bibfnamefont {V.~E.}\ \bibnamefont
  {Colussi}}, \bibinfo {author} {\bibfnamefont {J.~P.}\ \bibnamefont
  {Corson}},\ and\ \bibinfo {author} {\bibfnamefont {J.~P.}\ \bibnamefont
  {D’Incao}},\ }\bibfield  {title} {\bibinfo {title} {{Dynamics of three-body
  correlations in quenched unitary Bose gases}},\ }\href
  {https://doi.org/10.1103/PhysRevLett.120.100401} {\bibfield  {journal}
  {\bibinfo  {journal} {Phys.\ Rev.\ Lett.}\ }\textbf {\bibinfo {volume}
  {120}},\ \bibinfo {pages} {100401} (\bibinfo {year} {2018})}\BibitemShut
  {NoStop}%
\bibitem [{\citenamefont {Colussi}\ \emph {et~al.}(2019)\citenamefont
  {Colussi}, \citenamefont {van Zwol}, \citenamefont {D'Incao},\ and\
  \citenamefont {Kokkelmans}}]{colussi2019bunching}%
  \BibitemOpen
  \bibfield  {author} {\bibinfo {author} {\bibfnamefont {V.~E.}\ \bibnamefont
  {Colussi}}, \bibinfo {author} {\bibfnamefont {B.~E.}\ \bibnamefont {van
  Zwol}}, \bibinfo {author} {\bibfnamefont {J.~P.}\ \bibnamefont {D'Incao}},\
  and\ \bibinfo {author} {\bibfnamefont {S.~J. J. M.~F.}\ \bibnamefont
  {Kokkelmans}},\ }\bibfield  {title} {\bibinfo {title} {{Bunching, clustering,
  and the buildup of few-body correlations in a quenched unitary Bose gas}},\
  }\href {https://doi.org/10.1103/PhysRevA.99.043604} {\bibfield  {journal}
  {\bibinfo  {journal} {Phys.\ Rev.~A}\ }\textbf {\bibinfo {volume} {99}},\
  \bibinfo {pages} {043604} (\bibinfo {year} {2019})}\BibitemShut {NoStop}%
\bibitem [{\citenamefont {Bekassy}\ and\ \citenamefont
  {Hofmann}(2022)}]{bekassy2022}%
  \BibitemOpen
  \bibfield  {author} {\bibinfo {author} {\bibfnamefont {V.}~\bibnamefont
  {Bekassy}}\ and\ \bibinfo {author} {\bibfnamefont {J.}~\bibnamefont
  {Hofmann}},\ }\bibfield  {title} {\bibinfo {title} {{Nonrelativistic
  Conformal Invariance in Mesoscopic Two-Dimensional Fermi Gases}},\ }\href
  {https://doi.org/10.1103/PhysRevLett.128.193401} {\bibfield  {journal}
  {\bibinfo  {journal} {Phys.\ Rev.\ Lett.}\ }\textbf {\bibinfo {volume}
  {128}},\ \bibinfo {pages} {193401} (\bibinfo {year} {2022})}\BibitemShut
  {NoStop}%
\bibitem [{\citenamefont {Dusling}\ and\ \citenamefont
  {Sch{\"a}fer}(2013)}]{dusling2013}%
  \BibitemOpen
  \bibfield  {author} {\bibinfo {author} {\bibfnamefont {K.}~\bibnamefont
  {Dusling}}\ and\ \bibinfo {author} {\bibfnamefont {T.}~\bibnamefont
  {Sch{\"a}fer}},\ }\bibfield  {title} {\bibinfo {title} {{Bulk viscosity and
  conformal symmetry breaking in the dilute Fermi gas near unitarity}},\ }\href
  {https://doi.org/10.1103/PhysRevLett.111.120603} {\bibfield  {journal}
  {\bibinfo  {journal} {Phys.\ Rev.\ Lett.}\ }\textbf {\bibinfo {volume}
  {111}},\ \bibinfo {pages} {120603} (\bibinfo {year} {2013})}\BibitemShut
  {NoStop}%
\bibitem [{\citenamefont {Enss}(2019)}]{enss2019bulk}%
  \BibitemOpen
  \bibfield  {author} {\bibinfo {author} {\bibfnamefont {T.}~\bibnamefont
  {Enss}},\ }\bibfield  {title} {\bibinfo {title} {{Bulk Viscosity and Contact
  Correlations in Attractive Fermi Gases}},\ }\href
  {https://doi.org/10.1103/PhysRevLett.123.205301} {\bibfield  {journal}
  {\bibinfo  {journal} {Phys.\ Rev.\ Lett.}\ }\textbf {\bibinfo {volume}
  {123}},\ \bibinfo {pages} {205301} (\bibinfo {year} {2019})}\BibitemShut
  {NoStop}%
\bibitem [{\citenamefont {Nishida}(2019)}]{nishida2019}%
  \BibitemOpen
  \bibfield  {author} {\bibinfo {author} {\bibfnamefont {Y.}~\bibnamefont
  {Nishida}},\ }\bibfield  {title} {\bibinfo {title} {Viscosity spectral
  functions of resonating fermions in the quantum virial expansion},\ }\href
  {https://doi.org/10.1016/j.aop.2019.167949} {\bibfield  {journal} {\bibinfo
  {journal} {Ann.\ Phys.\ (N.Y.)}\ }\textbf {\bibinfo {volume} {410}},\
  \bibinfo {pages} {167949} (\bibinfo {year} {2019})}\BibitemShut {NoStop}%
\bibitem [{\citenamefont {Hofmann}(2020)}]{hofmann2020}%
  \BibitemOpen
  \bibfield  {author} {\bibinfo {author} {\bibfnamefont {J.}~\bibnamefont
  {Hofmann}},\ }\bibfield  {title} {\bibinfo {title} {High-temperature
  expansion of the viscosity in interacting quantum gases},\ }\href
  {https://doi.org/10.1103/PhysRevA.101.013620} {\bibfield  {journal} {\bibinfo
   {journal} {Phys.\ Rev.~A}\ }\textbf {\bibinfo {volume} {101}},\ \bibinfo
  {pages} {013620} (\bibinfo {year} {2020})}\BibitemShut {NoStop}%
\bibitem [{\citenamefont {Fujii}\ and\ \citenamefont
  {Nishida}(2020)}]{fujii2020}%
  \BibitemOpen
  \bibfield  {author} {\bibinfo {author} {\bibfnamefont {K.}~\bibnamefont
  {Fujii}}\ and\ \bibinfo {author} {\bibfnamefont {Y.}~\bibnamefont
  {Nishida}},\ }\bibfield  {title} {\bibinfo {title} {{Bulk viscosity of
  resonating fermions revisited: Kubo formula, sum rule, and the dimer and
  high-temperature limits}},\ }\href
  {https://doi.org/10.1103/PhysRevA.102.023310} {\bibfield  {journal} {\bibinfo
   {journal} {Phys.\ Rev.~A}\ }\textbf {\bibinfo {volume} {102}},\ \bibinfo
  {pages} {023310} (\bibinfo {year} {2020})}\BibitemShut {NoStop}%
\bibitem [{\citenamefont {Maki}\ and\ \citenamefont
  {Zhang}(2020)}]{maki2020role}%
  \BibitemOpen
  \bibfield  {author} {\bibinfo {author} {\bibfnamefont {J.}~\bibnamefont
  {Maki}}\ and\ \bibinfo {author} {\bibfnamefont {S.}~\bibnamefont {Zhang}},\
  }\bibfield  {title} {\bibinfo {title} {{Role of Effective Range in the Bulk
  Viscosity of Resonantly Interacting $s$- and $p$-Wave Fermi Gases}},\ }\href
  {https://doi.org/10.1103/PhysRevLett.125.240402} {\bibfield  {journal}
  {\bibinfo  {journal} {Phys.\ Rev.\ Lett.}\ }\textbf {\bibinfo {volume}
  {125}},\ \bibinfo {pages} {240402} (\bibinfo {year} {2020})}\BibitemShut
  {NoStop}%
\bibitem [{\citenamefont {Frank}\ \emph {et~al.}(2020)\citenamefont {Frank},
  \citenamefont {Zwerger},\ and\ \citenamefont {Enss}}]{frank2020quantum}%
  \BibitemOpen
  \bibfield  {author} {\bibinfo {author} {\bibfnamefont {B.}~\bibnamefont
  {Frank}}, \bibinfo {author} {\bibfnamefont {W.}~\bibnamefont {Zwerger}},\
  and\ \bibinfo {author} {\bibfnamefont {T.}~\bibnamefont {Enss}},\ }\bibfield
  {title} {\bibinfo {title} {{Quantum critical thermal transport in the unitary
  Fermi gas}},\ }\href {https://doi.org/10.1103/PhysRevResearch.2.023301}
  {\bibfield  {journal} {\bibinfo  {journal} {Phys.\ Rev.\ Res.}\ }\textbf
  {\bibinfo {volume} {2}},\ \bibinfo {pages} {023301} (\bibinfo {year}
  {2020})}\BibitemShut {NoStop}%
\bibitem [{\citenamefont {Musolino}\ \emph {et~al.}(2022)\citenamefont
  {Musolino}, \citenamefont {Kurkjian}, \citenamefont {Van~Regemortel},
  \citenamefont {Wouters}, \citenamefont {Kokkelmans},\ and\ \citenamefont
  {Colussi}}]{musolino2022bose}%
  \BibitemOpen
  \bibfield  {author} {\bibinfo {author} {\bibfnamefont {S.}~\bibnamefont
  {Musolino}}, \bibinfo {author} {\bibfnamefont {H.}~\bibnamefont {Kurkjian}},
  \bibinfo {author} {\bibfnamefont {M.}~\bibnamefont {Van~Regemortel}},
  \bibinfo {author} {\bibfnamefont {M.}~\bibnamefont {Wouters}}, \bibinfo
  {author} {\bibfnamefont {S.~J. J. M.~F.}\ \bibnamefont {Kokkelmans}},\ and\
  \bibinfo {author} {\bibfnamefont {V.~E.}\ \bibnamefont {Colussi}},\
  }\bibfield  {title} {\bibinfo {title} {{Bose-Einstein Condensation of
  Efimovian Triples in the Unitary Bose Gas}},\ }\href
  {https://doi.org/10.1103/PhysRevLett.128.020401} {\bibfield  {journal}
  {\bibinfo  {journal} {Phys.\ Rev.\ Lett.}\ }\textbf {\bibinfo {volume}
  {128}},\ \bibinfo {pages} {020401} (\bibinfo {year} {2022})}\BibitemShut
  {NoStop}%
\end{thebibliography}%

\end{document}